\documentclass[superscriptaddress, prd, aps,amsmath,amssymb,showpacs,showkeys, twocolumn]{revtex4-2}
\usepackage[dvips]{graphicx,color}
\usepackage{subfig}
\usepackage{times}
\usepackage{xcolor}
\usepackage[%
  colorlinks=true,
  urlcolor=blue,
  linkcolor=red,
  citecolor=blue
]{hyperref}
\usepackage{orcidlink}

\begin{document}
\title{Quasinormal Modes and Optical Properties of 4-D black holes in Einstein Power-Yang-Mills Gravity}

\author{Dhruba Jyoti Gogoi\orcidlink{0000-0002-4776-8506}}
\email[Email: ]{moloydhruba@yahoo.in}

\affiliation{Department of Physics, Dibrugarh University,
Dibrugarh 786004, Assam, India.}

\author{Jyatsnasree Bora \orcidlink{0000-0001-9751-5614}}
\email[Email: ]{jyatnasree.borah@gmail.com}

\affiliation{Department of Physics, Dibrugarh University,
Dibrugarh 786004, Assam, India.}


\author{M. Koussour\orcidlink{0000-0002-4188-0572}}
\email[Email: ]{pr.mouhssine@gmail.com}
\affiliation{Quantum Physics and Magnetism Team, LPMC, Faculty of Science Ben
M'sik,\\
Casablanca Hassan II University,
Morocco.}

\author{Yassine Sekhmani \orcidlink{0000-0001-7448-4579}}
\email[Email: ]{sekhmaniyassine@gmail.com}

\affiliation{D\'epartement de Physique, Equipe des Sciences de la mati\'ere et du rayonnement, ESMaR, Facult\'e des Sciences, Universit\'e Mohammed V de Rabat, Rabat, Morocco.}

\begin{abstract}

This paper explores the impact of the Yang-Mills charge parameter and the exponent term on a $4$D black hole solution in the Einstein Power-Yang-Mills theory. Through an investigation of the massless scalar quasinormal mode spectrum, black hole shadow, and emission rate, we have determined that the effects of these two parameters are opposite. Specifically, the Yang-Mills charge parameter causes an increase in the real quasinormal frequencies with a correspondingly smaller damping rate. It also results in a smaller black hole shadow and a lower evaporation rate.
\end{abstract}


\maketitle
\section{Introduction}\label{sec01}

Black holes, mesmerizing celestial entities subject to the principles of Einstein's theory of gravitation, hold a prominent position in the cosmos. A groundbreaking milestone in black hole research was achieved with the detection of Gravitational Waves (GWs) on September 14th, 2015 \cite{PhysRevLett.116.061102}. This discovery not only advanced our understanding of black holes but also provided a novel avenue for experimentally testing theories of gravity. According to Einstein's theory of general relativity, GWs are predicted to arise from the acceleration of massive objects, generating ripples in the fabric of spacetime. These waves carry vital information about the kinematics and dynamics of the astrophysical sources responsible for their generation. Advanced instruments such as LIGO and Virgo are capable of detecting GWs.

During the merger of two black holes, a final black hole is formed, which emits GWs exhibiting unique wave patterns known as ring-down modes. These GWs possess quasinormal modes that depend on the mass and spin of the resulting black hole. Analyzing GW data through these quasinormal modes is pivotal in unraveling the enigmatic properties of black holes and gaining valuable insights into their nature.

Quasinormal modes are a significant and intriguing characteristic of black hole physics. They are oscillations of a black hole that gradually weaken over time and are distinguished by intricate frequencies. Quasinormal modes are termed ``quasinormal" because they are not precisely normal modes, which would continue to oscillate infinitely \cite{Vishveshwara:1970zz, Press:1971wr,Kokkotas:1999bd}. Rather, they fade away due to dissipative mechanisms like the emission of gravitational waves. Quasinormal modes are complex values that represent the emission of gravitational waves from compact and massive objects in the Universe. The emission frequency is denoted by the real component of the quasinormal modes, while the imaginary component corresponds to its decay. Understanding quasinormal modes is crucial because they contain information about the black hole's characteristics, such as its mass, angular momentum, and the properties of the surrounding spacetime. Furthermore, investigating quasinormal modes provides insights into the nature of black holes and the strong gravity regime, which is challenging to explore using other methods. These modes are fundamental to understanding the structure and evolution of black holes and their role in astrophysical phenomena such as gravitational wave signals. In recent years, the exploration of GWs and the quasinormal modes exhibited by black holes has witnessed extensive investigation within various modified gravity theories \cite{Rincon:2018ktz, Liu:2022ygf, Rincon:2021gwd, Ovgun:2018gwt, Ovgun:2019yor, Anacleto:2021qoe, Lambiase:2023hng, sekhmani_electromagnetic_2023, Gogoi:2023kjt, Parbin:2022iwt, karmakar_quasinormal_2022,Gogoi:2022wyv,Gogoi:2021cbp,Gogoi:2021dkr,Pantig:2022gih}.

In recent times, there has been a surge of interest in gravitational theories that incorporate the nonlinearity of Maxwell fields. The scientific community has become intrigued by the nonsingular nature of black hole solutions found in nonlinear electrodynamics \cite{Ayon-Beato:1998hmi}. Furthermore, a set of black hole solutions has been derived within the framework of power Maxwell invariant theory, where the Lagrangian density is expressed as $(\mathcal{F}_{\mu\nu} \mathcal{F}^{\mu\nu})^q$, with $q$ representing an arbitrary rational number \cite{Maeda:2008ha}. Having examined black hole solutions within the context of Einstein power Maxwell invariant gravity, researchers have delved into exploring other nonlinear models that involve the coupling of nonabelian Yang-Mills fields with gravity in the realm of general relativity.

For instance, the authors in \cite{Mazharimousavi:2009mb} investigated possible black hole solutions that are sourced by the power of Yang-Mills invariant, expressed as $(\mathcal{F}_{\alpha\beta}^{(a)} \mathcal{F}^{(a)\alpha\beta})^q$, where $\mathcal{F}_{\alpha\beta}^{(a)}$ is the Yang-Mills field with its internal index $1 \le a \le {\frac{1}{2}} (D-1) (D-2)$, and setting $q = 1$ recovers the $D$ dimensional Einstein-Yang-Mills black holes in $AdS$ space-time \cite{HabibMazharimousavi:2007fst,Mazharimousavi:2008ap}.
Such extensions studied in the context of black hole solutions have been found to exhibit new and interesting phenomena, such as novel thermodynamic properties. Despite the complexity involved in studying these theories, they remain important areas of research for physicists interested in understanding the fundamental nature of gravity and its interaction with other forces in the Universe.
Furthermore, the study of Van der Waals-like phase transitions and critical behaviour in the extended thermodynamics of $AdS$ black holes in Einstein Power-Maxwell and Einstein Power-Yang-Mills theories has been investigated \cite{Hendi:2017lgb, Zhang:2014eap, Yerra:2018mni,Hendi:2017mgb, Hendi:2016usw}. Recently, the Joule-Thomson expansion for higher dimensional non-linearly charged AdS black holes in Einstein-power Maxwell invariant gravity has been explored \cite{Feng:2020swq}. In another study, Joule-Thomson expansion of non-linearly charged adS black holes in Einstein-power-Yang-Mills gravity has been investigated in $D$ dimensions \cite{Biswas:2021uop}. However, the non-linearity of Yang-Mills theory naturally adds further complexity to the already non-linear nature of gravity. As a result, the theory and its associated solutions become quite intricate and difficult to analyze. This complexity presents a significant challenge for physicists trying to understand the properties and behaviour of these black holes. 

The defining characteristic of a black hole is its event horizon, which is the point beyond which no particles can escape due to the immense gravitational pull. This gravitational force traps all physical particles, including light, inside the event horizon, while outside it, light can escape \cite{Synge:1966okc}. The matter that surrounds a black hole and is pulled inward is known as accretion. As the accretion becomes heated due to viscous dissipation, it emits bright radiation at various frequencies, including radio waves that are detectable with radio telescopes. This creates a bright background with a dark area over it, known as the black hole shadow \cite{Luminet:1979nyg}. The idea of imaging the black hole shadow at the centre of our Milky Way was first proposed by Falcke et al. \cite{Falcke:1999pj}, although the concept has been around since the 1970s. Recently, the Event Horizon Telescope captured images of the black hole shadow in the Messier 87 galaxy and Sagittarius A* \cite{EventHorizonTelescope:2019dse,EventHorizonTelescope:2022xnr}. As a result, the black hole shadow has become a popular topic in current literature since the presence of a family of rotating black holes could provide information about the distortion of photon spheres due to the fact that rotating black holes have non-spherical horizons \cite{Ovgun:2018tua, EventHorizonTelescope:2021dqv, Belhaj:2020okh,Belhaj:2020rdb, Belhaj:2022kek, gogoi_joulethomson_2023, Ovgun:2020gjz,Ovgun:2019jdo,Kuang:2022xjp,Kumaran:2022soh,Mustafa:2022xod,Cimdiker:2021cpz,Okyay:2021nnh,Atamurotov:2022knb,Pantig:2022qak,Abdikamalov:2019ztb,Abdujabbarov:2016efm,Atamurotov:2015nra,Papnoi:2014aaa,Abdujabbarov:2012bn,Atamurotov:2013sca,Cunha:2018acu,Gralla:2019xty}. This results study of shadows in many kinds of modified gravities and also involves a comparative study with the observations of M87* and Sagittarius A* images in relation to size and shape geometric data \cite{Pantig:2022gih,Lobos:2022jsz, gogoi_joulethomson_2023, Uniyal:2023inx,Panotopoulos:2021tkk,Panotopoulos:2022bky,Khodadi:2022pqh,Khodadi:2021gbc,Zhao:2023uam,Khodadi:2020jij,Khodadi:2020gns,Vagnozzi:2022moj,Khodadi:2022ulo}. The shadow process plays a crucial role in the discovery of the thermodynamic phase transition, which leads to a special correspondence between the two processes, from which the shadow radius is considered an efficient tool to investigate thermodynamic black hole systems \cite{Zhang:2019glo}. Moreover, the shadow and quasi-normal modes relation is constructed in the eikonal limit, revealing that the real parts of the quasinormal modes are related to the shadow radii of black holes \cite{Jusufi:2020dhz,Jusufi:2020mmy}. 

Motivated by these studies, in this work, we intend to consider $4$-D Einstein-power-Yang-Mills gravity to investigate the quasinormal mode spectrum and optical properties of the black hole.

The paper is organised as follows. In Section \ref{sec02}, we discuss the black hole solution in Einstein Power- Yang-Mills gravity. Section \ref{sec03} deals with the study of quasinormal modes. In Section \ref{sec04}, optical behaviours of the black hole {\it viz.}, shadow and emission rate have been discussed. In Section \ref{sec05},  we discuss the connection of quasinormal modes and the optical properties in brief. Finally, in Section \ref{sec06}, concluding remarks are given.

Throughout the manuscript, we consider $c=\hbar = 8\pi G = 1$.

\section{4-D Black Hole solution in Einstein Power-Yang-Mills Gravity}
\label{sec02}

Initially, we begin with the $D$ dimensional action for Einstein-power-Yang-Mills gravity with non-vanishing cosmological constant given by \cite{Mazharimousavi:2009mb}
\begin{equation}
I = \frac{1}{2}\int d^Dx \sqrt{-g}\Big (\mathcal{R} - \frac {(D-2)(D-1)}{3}\Lambda - \mathcal{F}^q\Big ),
\label{action11}
\end{equation}
where $R$ is the Ricci Scalar, $q$ is a positive real parameter and $\mathbf{F}$ is the Yang-Mills invariant, expressed as
\begin{eqnarray}
\mathbf{F} &=& Tr(\mathcal{F}_{\lambda\sigma}^{(a)} \mathcal{F}^{(a)\lambda\sigma}), \nonumber \\
Tr(.) &= &\Sigma_{a=1}^{\frac {(D-1)(D-2)}{2}} (.).
\label{action22}
\end{eqnarray}
The Yang-Mills field is defined as \cite{Biswas:2022qyl}
\begin{equation}
\mathcal{F}_{\lambda\sigma}^{(a)}=\partial_\mu A_\nu^{(a)} - \partial_\nu A_\mu^{(a)} + \frac{1}{2\sigma} C_{(b)(c)}^{(a)}A_\mu^{(b)}A_\mu^{(c)}.
\label{YM}
\end{equation}
Here $C_{(b)(c)}^{(a)}$ are the structure constants of $\frac {(D-1)(D-2)}{2}$ parameter Lie group $G$ and $\sigma$ is a coupling constant, $A_\mu^{(a)}$ are the $SO(D-1)$ gauge group Yang-Mills potentials. We have considered the following metric ansatz for $D$ dimensions
\begin{equation}
ds^2=-f(r) dt^2+\frac{dr^2}{f(r)}+r^2 d\Omega^2_{(D-2)}.
\end{equation}
Here $d\Omega^2_{D-2}$ stands for the line element associated with unit $(D-2)$ sphere. Assuming $q\neq \frac{(D-1)}{4}$, and in the presence of cosmological constant we have the metric function of the Einstein-power-Yang-Mills black hole as given by \cite{Biswas:2022qyl}
\begin{equation}
f(r)=1-\frac{2M}{r^{D-3}}-\frac{\Lambda}{3} r^2+\frac{[(D-3)(D-2)Q^2]^q}{(D-2)(4q-D+1) r^{4q-2}}.
\label{EPYM}
\end{equation}
The parameter $M$ corresponds to the mass of the black hole, whereas $Q$ represents a charge parameter associated with Yang-Mills fields.
{ For $D = 4$, the metric function in Eq. \eqref{EPYM} can be expressed in the following manner} \cite{Biswas:2022qyl}:
\begin{equation}
    f(r) = 1-\frac{2 M}{r}+\frac{2^{q-1} Q^{2q} r^{2-4 q}}{4 q-3}-\frac{\Lambda  r^2}{3}.
\end{equation}
In this study, we shall use this metric function to investigate the quasinormal modes and optical properties of the black hole. In order to satisfy the Weak Energy Condition (WEC) of the Power-Yang-Mills term, a positive value of $q$ is necessary \cite{Mazharimousavi:2009mb}. To satisfy various energy conditions, including the causality condition, a limit for $q$ has been discussed in \cite{Mazharimousavi:2009mb}, given by $\frac{3}{4}\le q < \frac{3}{2}$. When $q=1$, the aforementioned black hole solutions are simplified into Einstein-Yang-Mills black holes \cite{HabibMazharimousavi:2007fst, Mazharimousavi:2008ap}.
\section{Quasinormal modes}
\label{sec03}

In this section, we will address the massless scalar perturbation 
in the spacetime of a black hole. We will assume that the test 
field exerts negligible influence on the black hole spacetime. To 
determine the quasinormal modes, we will derive Schr\"odinger-like 
wave equations which should be of Klein-Gordon type for the case, 
taking into account the corresponding conservation relations of the 
concerned spacetime. { In this study, we shall use the Pad\'e 
averaged 6th order WKB approximation method, to calculate the quasinormal modes. }

Considering only axial perturbation, we can express the perturbed metric as presented in \cite{Bouhmadi-Lopez:2020oia, Gogoi:2023kjt}:
\begin{align} \label{pert_metric}
ds^2 =& -\, |g_{tt}|\, dt^2 + r^2 \sin^2\!\theta\, (d\phi - p_1(t,r,\theta)\,
dt \notag \\ &- p_2(t,r,\theta)\, dr - p_3(t,r,\theta)\, d\theta)^2 + g_{rr}\, dr^2 \notag \\ & +
r^2 d\theta^2,
\end{align}
where the parameters $p_1$, $p_2$, and $p_3$ define the 
perturbation introduced to the black hole spacetime. The metric 
functions $g_{tt}$ and $g_{rr}$ represent the zeroth order terms.
\subsection{Scalar Perturbation}
First, we shall consider the massless scalar field near the 
previously established black hole. As it is considered that the 
effect of scalar field on the spacetime is minimal, the perturbed 
metric Eq. \eqref{pert_metric} can be expressed as follows: 
\begin{equation}
ds^2 = -\,|g_{tt}|\, dt^2 + g_{rr}\, dr^2 +r^2 d \Omega^2.
\end{equation}
Now, for this case, it's feasible to write the Klein-Gordon equation in curved spacetime as
\begin{equation}  \label{scalar_KG}
\square \Phi = \dfrac{1}{\sqrt{-g}} \partial_\mu (\sqrt{-g} g^{\mu\nu}
\partial_\nu \Phi) = 0.
\end{equation}
With the help of this equation Eq. \eqref{scalar_KG} the quasinormal 
modes associated with the scalar perturbation can be described. Now 
the scalar field can be decomposed as follows:
\begin{equation}\label{scalar_field}
\Phi(t,r,\theta, \phi) = \dfrac{1}{r} \sum_{l,m} \psi_l(t,r) Y_{lm}(\theta,
\phi).
\end{equation}
In this equation, we have used spherical harmonics and $l$ and $m$ 
are the indices associated with it. Also, $\psi_l(t,r)$ is the 
radial time-dependent wave function. Using Eq. \eqref{scalar_KG} 
and \eqref{scalar_field} one can have,
\begin{equation}  \label{radial_scalar}
\partial^2_{r_*} \psi(r_*)_l + \omega^2 \psi(r_*)_l = V_s(r) \psi(r_*)_l.
\end{equation}
Here in this expression, $r_*$ is given as 
\begin{equation}  \label{tortoise}
\dfrac{dr_*}{dr} = \sqrt{g_{rr}\, |g_{tt}^{-1}|}
\end{equation}
and is known as the tortoise coordinate. The term $V_s(r)$ 
represents the effective potential with its explicit form as given 
below: 
\begin{equation}  \label{Vs}
V_s(r) = |g_{tt}| \left( \dfrac{l(l+1)}{r^2} +\dfrac{1}{r \sqrt{|g_{tt}|
g_{rr}}} \dfrac{d}{dr}\sqrt{|g_{tt}| g_{rr}^{-1}} \right),
\end{equation}
here, the term $l$ represents the multipole moment of the black hole's
quasinormal modes.

  \begin{figure}[ht!]
      	\centering{
      	\includegraphics[scale=0.8]{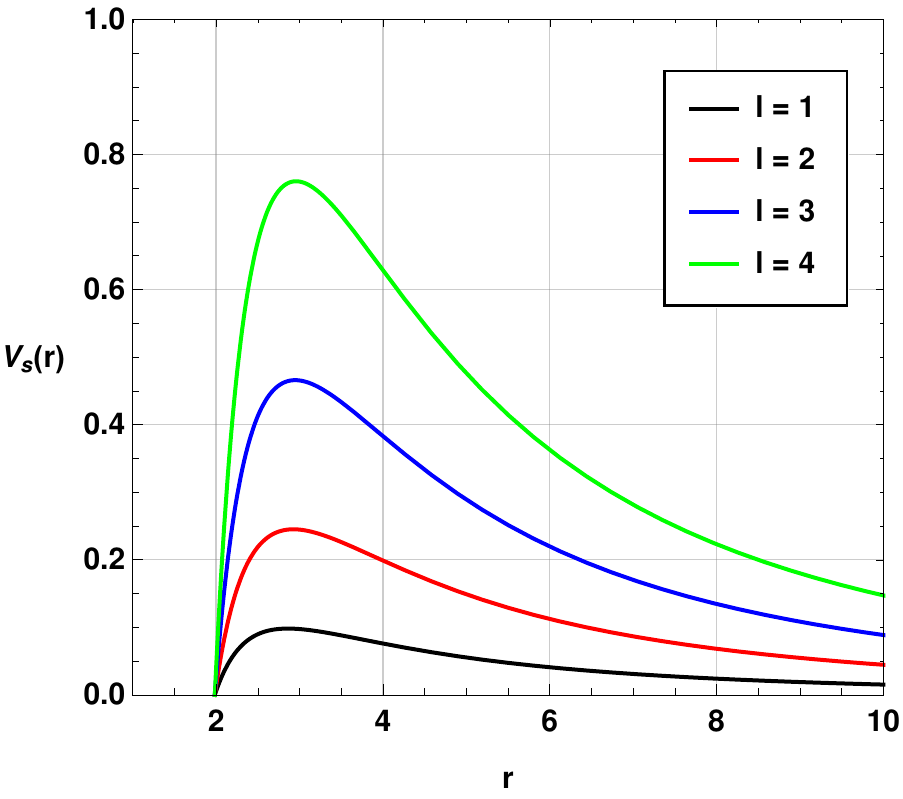}}
      	\caption{Variation scalar potential with respect to $r$ using $M = 1$, $Q = 0.3$, $\Lambda = 0.002$ and $q = 1.1$.}
      	\label{figPot01}
      \end{figure}
      \begin{figure*}[t!]
      	\centering{
      	\includegraphics[scale=0.8]{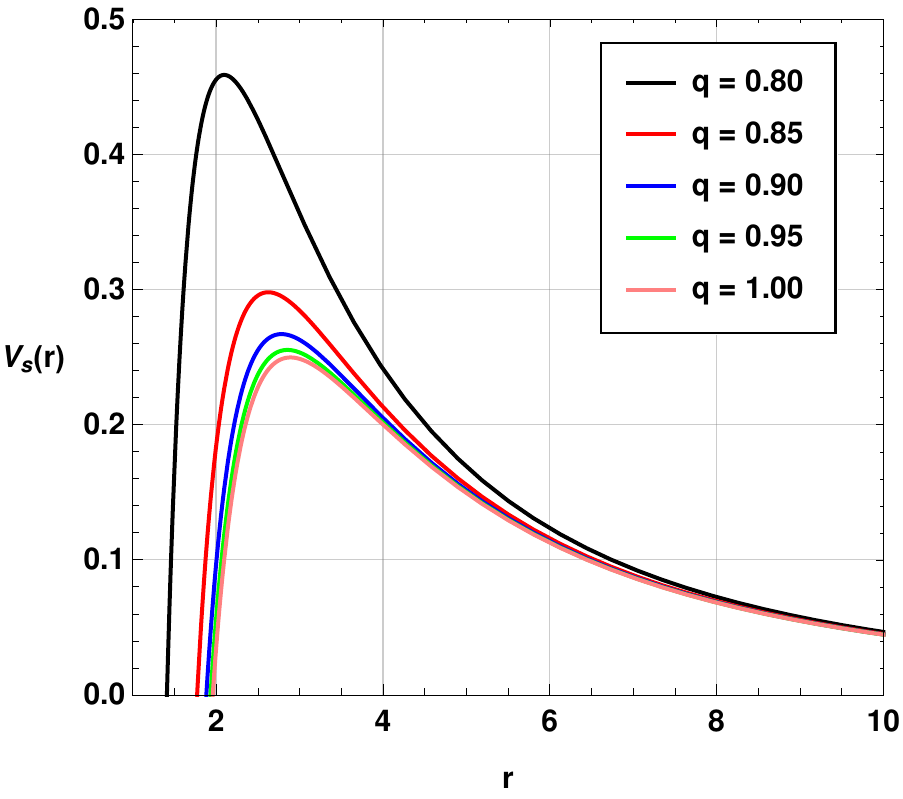} \hspace{1cm}
       \includegraphics[scale=0.8]{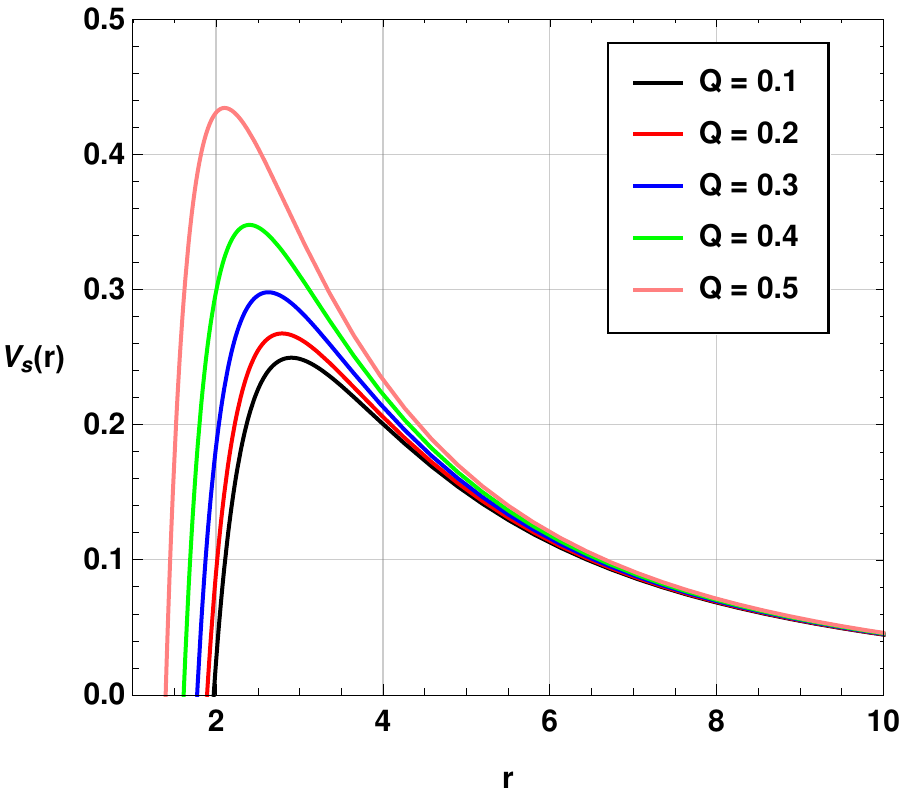}}
      	\caption{Variation scalar potential with respect to $r$ using $M = 1$ and $\Lambda = 0.002$. On the first panel, we have used $Q = 0.3$ and on the second $q = 0.85$.}
      	\label{figPot02}
      \end{figure*}  
In Fig. \ref{figPot01} and \ref{figPot02}, we have shown the behaviour of the scalar potential for different values of the model parameters. In Fig. \ref{figPot01}, we show the impacts of multipole moment $l$ on the behaviour of potential. In the first panel of Fig. \ref{figPot02}, we show the behaviour of the potential for different values of the exponent term $q$. It is seen that for the lower values of $q$, the peak value of potential increases drastically and the peak moves towards $r=0$. However, for higher values of $q$ near $1$, the variation of the potential is small and the peak also decreases slowly. The peak moves away from $r = 0$ as $q$ increases. In the second panel of Fig. \ref{figPot02}, we show how $Q$ impacts the potential of the scalar perturbation. It is seen that with an increase in the value of $Q$, the peak value of the potential increases non-linearly and the peak moves towards $r = 0$. Thus the behaviour of the potential depicts that quasinormal modes might change non-linearly with respect to the model parameters $q$ and $Q$. We confirm this in the explicit calculation of quasinormal modes in the next part of our study.

\subsection{The Pad\'e averaged WKB approximation method}
This investigation employed the Pad\'e averaged sixth-order WKB approximation technique. Within this method, the oscillation frequency $\omega$ of GWs can be determined using the following expression:
\begin{equation}
\omega = \sqrt{-\, i \left[ (n + 1/2) + \sum_{k=2}^6 \bar{\Lambda}_k \right] \sqrt{-2 V_0''} + V_0},
\end{equation}
where $n = 0, 1, 2\hdots$, stands for overtone number, $V_0 = V|_{r\, =\, r_{max}}$ and 
$V_0'' = \dfrac{d^2 V}{dr^2}|_{r\, =\, r_{max}}$. The position $r_{max}$ corresponds to the location where the potential function $V(r)$ attains its highest value. The correction terms $\bar{\Lambda}_k$ play a role in refining the accuracy of the calculations. The specific expressions for these correction terms, along with the Pad\'e averaging method, can be found in the
Ref.s \cite{Schutz:1985km,Iyer:1986np,Konoplya:2003ii,Matyjasek:2019eeu,Konoplya:2019hlu}.
These references provide detailed information regarding the mathematical forms of the correction terms and the procedure for Pad\'e averaging.
\begin{widetext}

\begin{table}[htb!]
{\centering
\begin{center}
\caption{The quasinormal modes of the black hole have been computed for varying values of the multipole moment $l$ and an overtone number of $n=0$ using the 6th order Pad\'e averaged WKB approximation method. The specific values assigned to the model parameters in this calculation are $M=1$, $Q=0.2$, $q=0.9$, and $\Lambda=0.002$. }
\vspace{0.3cm}
\begin{tabular}{cccc}
\hline\\[-10pt]
$l$ & Pad\'e averaged WKB & $\Delta_{rms}$ & $\Delta_6$ \\[2pt]  \hline
\\[-10pt] 

 $1$ & $0.296633\, -0.098464 i$ & $0.000025161$ & $6.342137\times 10^{-6}$ \\
 $2$ & $0.490184\, -0.0974166 i$ & $1.816291\times 10^{-6}$ & $6.036433\times 10^{-7}$ \\
 $3$ & $0.684718\, -0.0971061 i$ & $1.226110\times 10^{-6}$ & $1.074288\times 10^{-7}$ \\
 $4$ & $0.879551\, -0.0969766 i$ & $1.028066\times 10^{-6}$ & $2.924839\times 10^{-8}$ \\
 $5$ & $1.07451\, -0.0969103 i$ & $4.828103\times 10^{-7}$ & $4.094713\times 10^{-8}$ \\
 $6$ & $1.26955\, -0.0968721 i$ & $1.418885\times 10^{-7}$ & $1.678297\times 10^{-8}$ \\
 $7$ & $1.46462\, -0.096848 i$ & $5.412103\times 10^{-9}$ & $2.204433\times 10^{-9}$ \\
 $8$ & $1.65972\, -0.0968318 i$ & $7.714701\times 10^{-9}$ & $2.274291\times 10^{-9}$ \\
 \hline  
\end{tabular}
\label{Table01}
\end{center}}
\end{table}
\end{widetext}
Table \ref{Table01} displays the quasinormal modes for various values of the multipole moment, specifically focusing on the case where the overtone number is $n=0$. The second column of the table presents the quasinormal modes obtained through the application of the 6th-order Pad\'e averaged WKB approximation method. Within this table, the term $\Delta_{rms}$ signifies the root mean square error associated with the 6th-order Pad\'e averaged WKB approximation method. Furthermore, the term $\Delta_6$ serves as a term to quantify the error between two adjacent order approximations. It is calculated by evaluating the absolute difference between $\omega_7$ and $\omega_5$ and then dividing the result by 2, as expressed by the following equation \cite{Konoplya:2003ii}:
\begin{equation}
\Delta_6 = \dfrac{|\omega_7 - \omega_5|}{2}.
\end{equation}
The quasinormal modes denoted as $\omega_7$ and $\omega_5$, are computed using the Pad\'e averaged 7th order and 5th order WKB approximation methods, respectively. It is noteworthy that as the multipole moment $l$ increases, the error associated with the quasinormal modes decreases. This behaviour is a characteristic pattern of the WKB approximation method, which demonstrates limitations in providing precise results when the overtone number $n$ exceeds the multipole moment $l$\cite{Lambiase:2023hng, sekhmani_electromagnetic_2023, Gogoi:2023kjt, Parbin:2022iwt, Gogoi:2022ove, karmakar_quasinormal_2022, Gogoi:2022wyv}.

\begin{figure*}[t!]
      	\centering{
      	\includegraphics[scale=0.55]{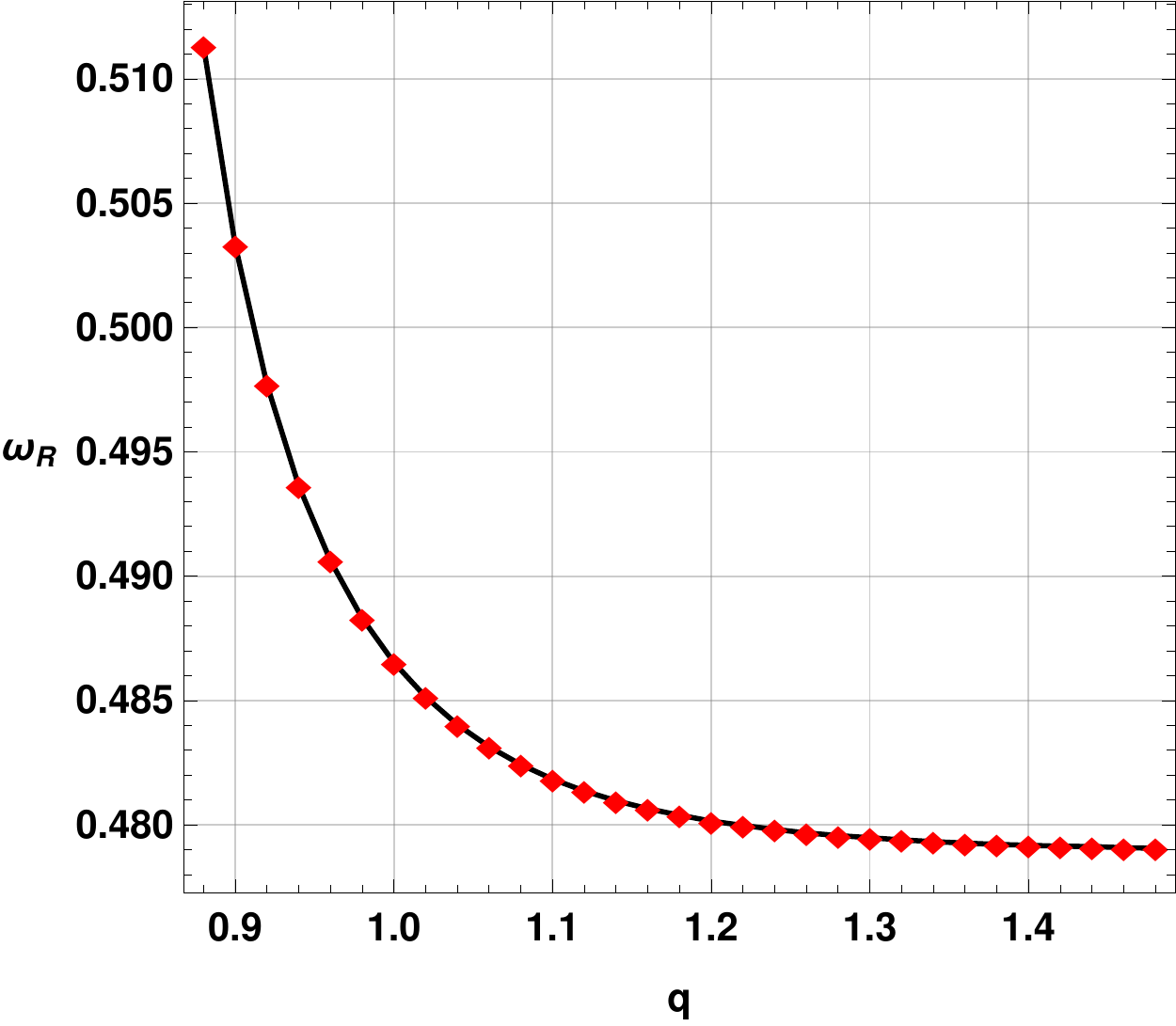}
       \includegraphics[scale=0.55]{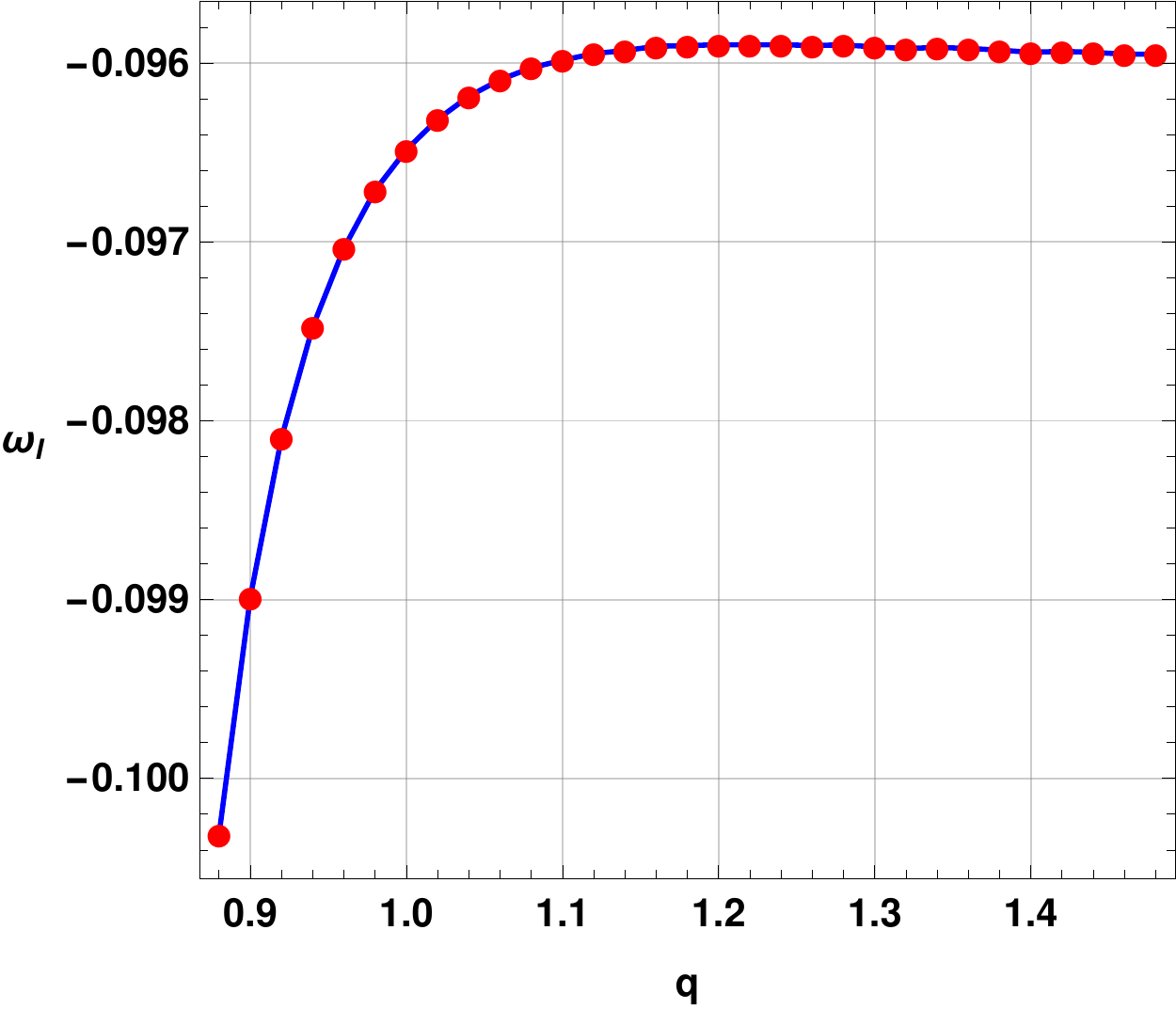}}
      	\caption{Variation of real and imaginary quasinormal modes using $M = 1$, $n=0$, $l = 2$, $Q = 0.3$ and $\Lambda = 0.002$. }
      	\label{figQNM01}
      \end{figure*}

      \begin{figure*}[t!]
      	\centering{
      	\includegraphics[scale=0.55]{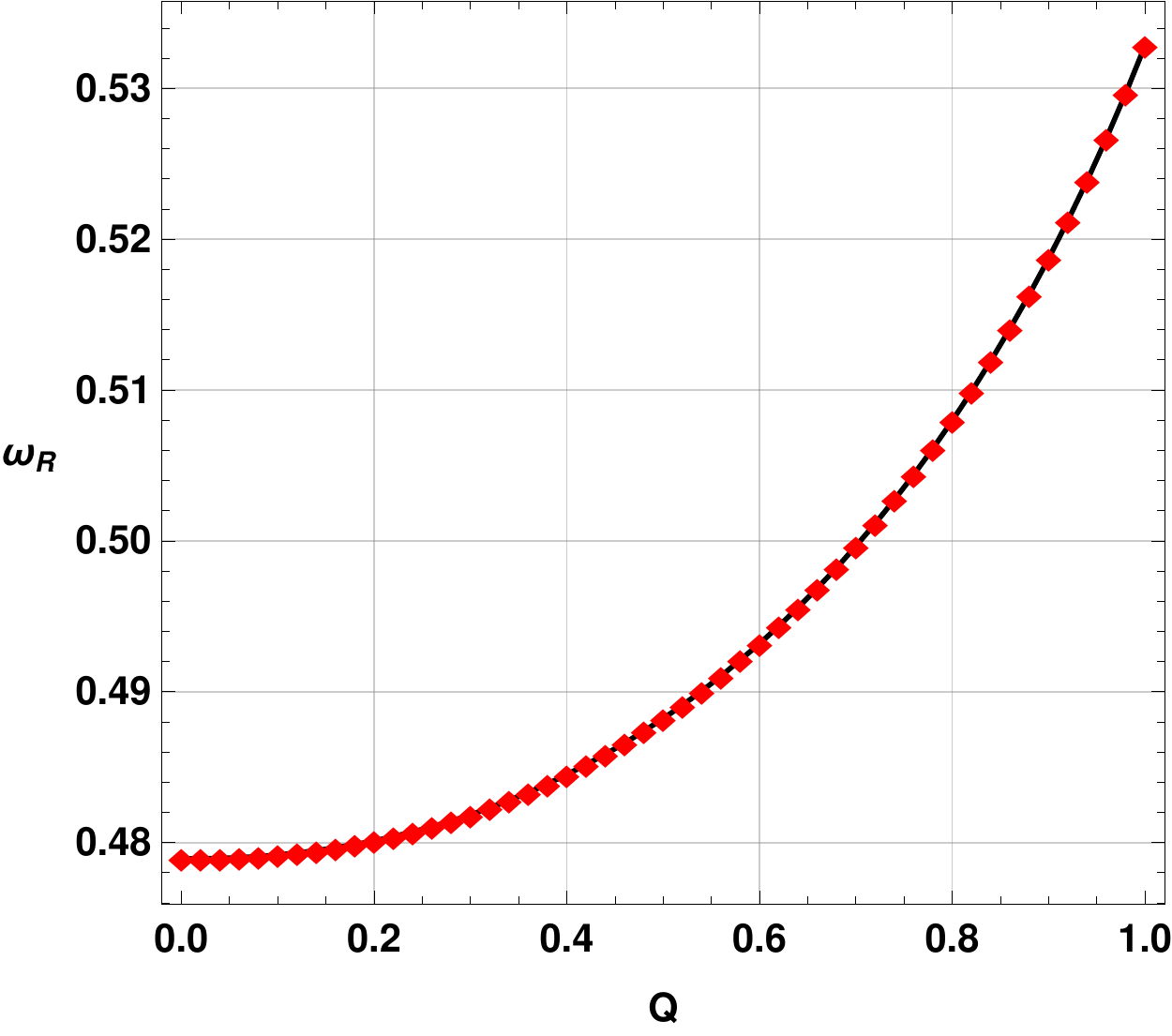}
       \includegraphics[scale=0.56]{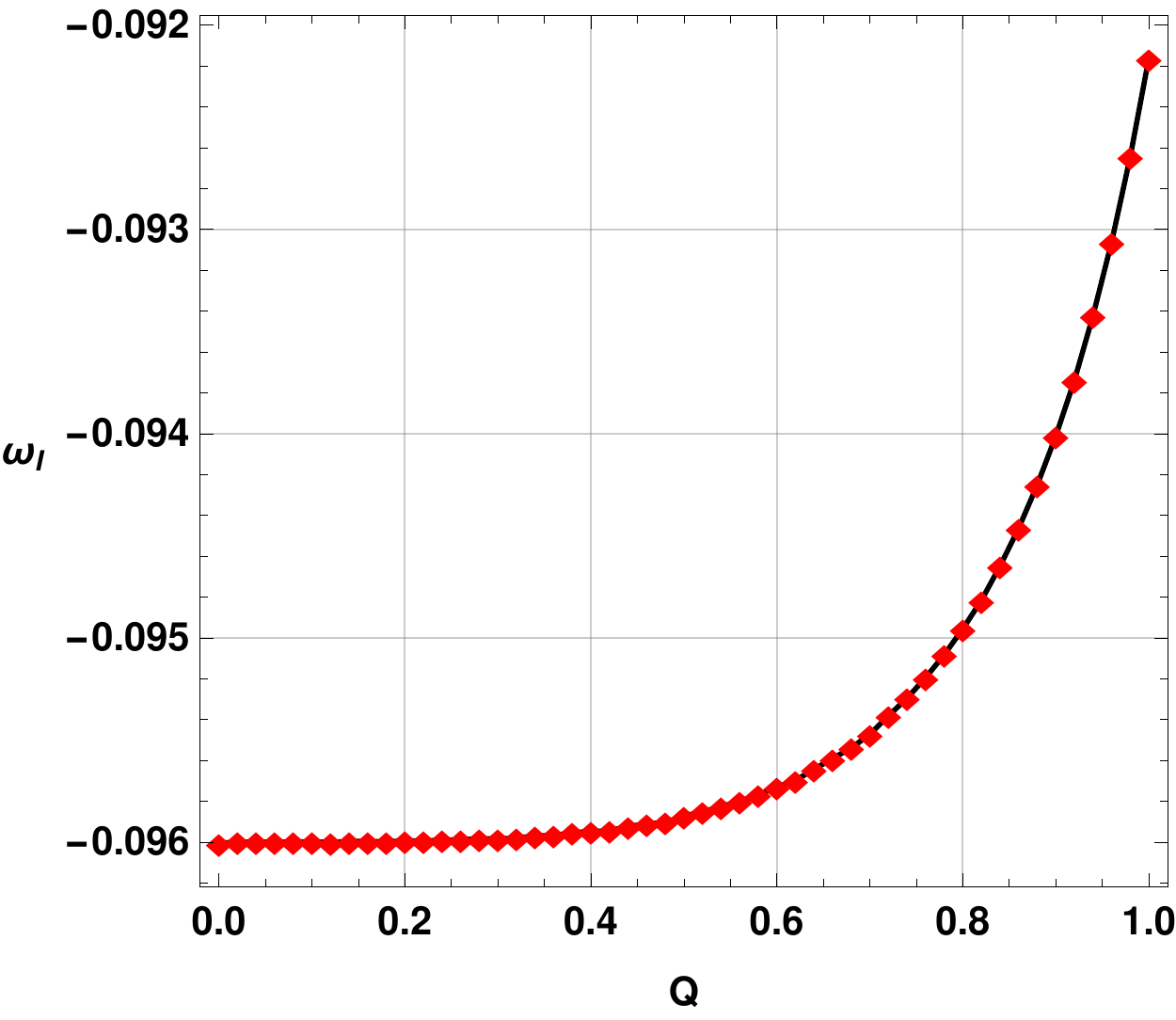}}
      	\caption{Variation of real and imaginary quasinormal modes using $M = 1$, $n=0$, $l = 2$, $q = 1.1$ and $\Lambda = 0.002$.}
      	\label{figQNM02}
      \end{figure*}

      \begin{figure*}[t!]
      	\centering{
      	\includegraphics[scale=0.55]{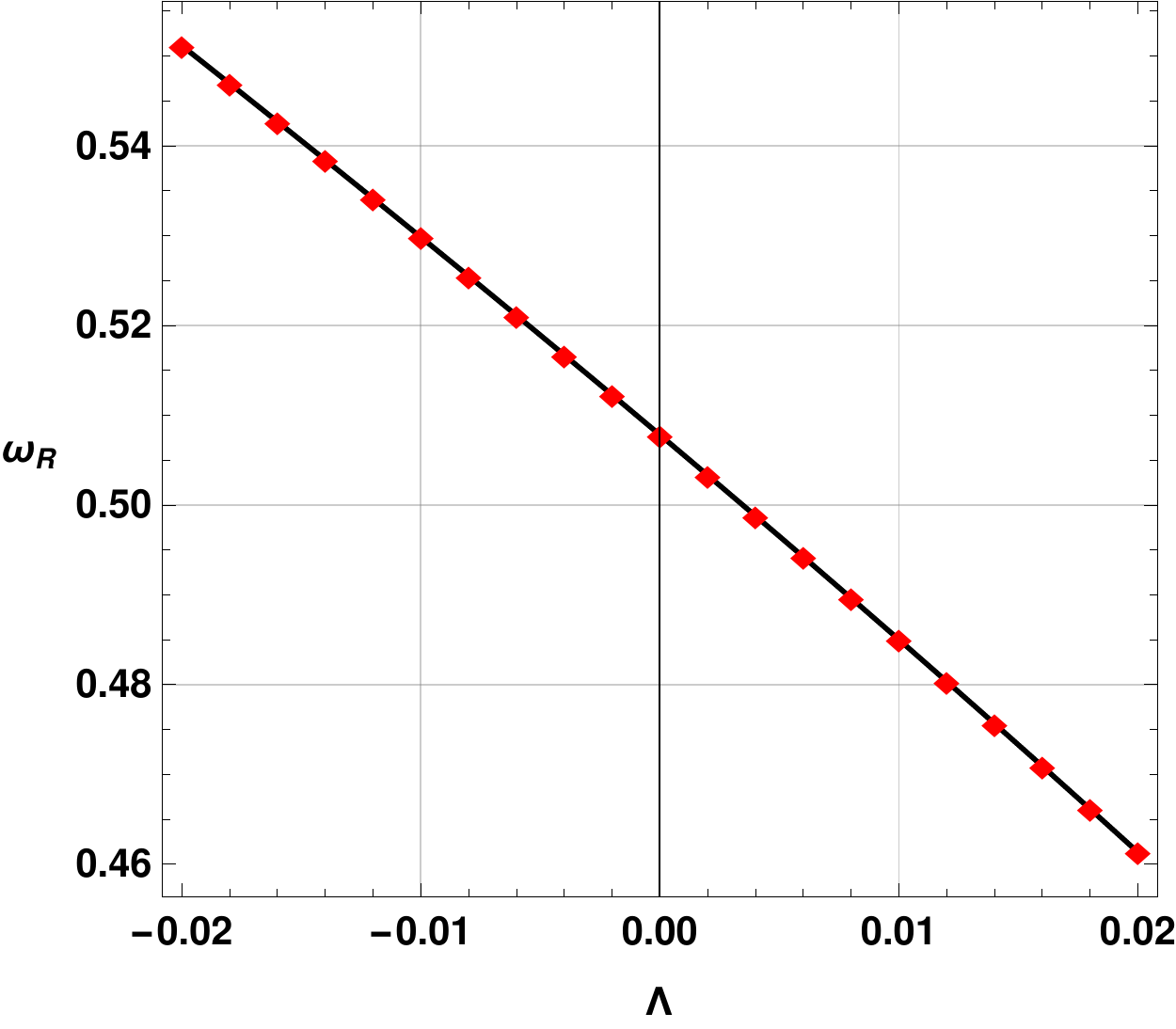}
       \includegraphics[scale=0.56]{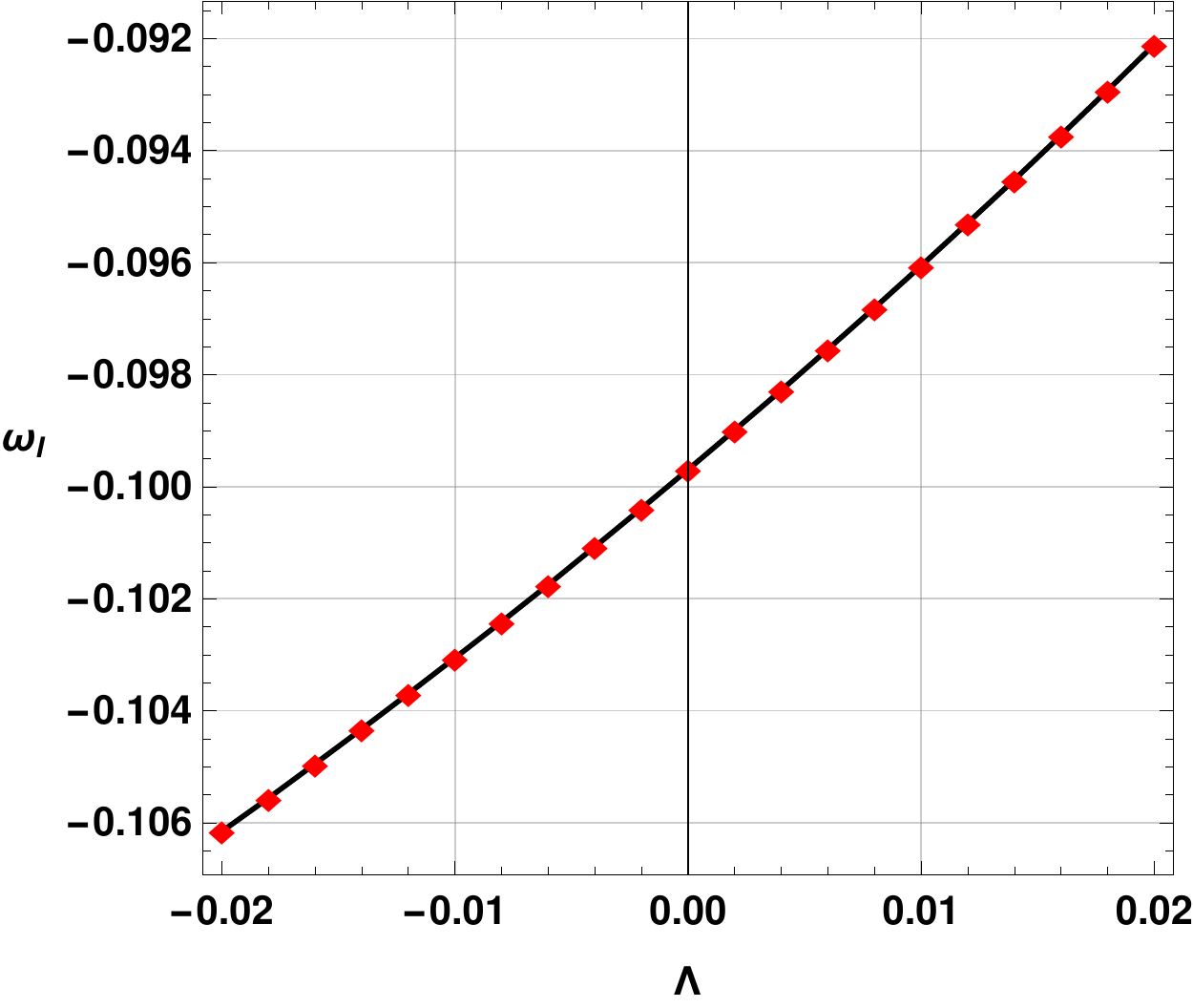}}
      	\caption{{ Variation of real and imaginary quasinormal modes using $M = 1$, $n=0$, $l = 2$, $q = 0.9$ and $Q = 0.3$.}}
      	\label{figQNM03}
      \end{figure*}

We have shown the variation of the massless scalar quasinormal modes in Fig. \ref{figQNM01} and \ref{figQNM02} with respect to $q$ and $Q$, respectively. Notably, when $q=1$, the theory converges to the conventional Einstein Yang-Mills theory. For values of $q$ less than $1$, the real quasinormal frequencies and damping rate both increase drastically (see Fig. \ref{figQNM01}). Conversely, for $q>1$, the real quasinormal frequencies slowly decrease with only minor variation. At approximately $q=1.5$, the real quasinormal frequencies approach a constant value. In terms of the damping rate, we observe a decrease with a minimum decay rate occurring at $q=1.2$. As $q$ surpasses this point, the damping rate slowly begins to increase. However, the damping rate becomes almost constant as $q$ approaches $1.5$.

In Fig. \ref{figQNM02}, on the first panel, we have shown the variation of the real quasinormal frequencies with respect to the Yang-Mills charge parameter $Q$. On the second panel, we have shown the variation of the damping rate of ringdown gravitational waves with respect to $Q$. It is clearly visible that as the Yang-Mills charge parameter $Q$ increases, the real quasinormal frequencies increase exponentially. On the other hand, with an increase in the value of $Q$, the damping rate decreases exponentially. Initially, up to $Q=0.4$, the decrease in the damping rate is almost negligible. However, beyond this point, the damping rate starts to decrease drastically, reaching a minimum value at $Q=1$.
Variation of quasinormal modes with respect to $\Lambda$ is shown in Fig. \ref{figQNM03}.

Hence, from the perspective of quasinormal modes, the impacts of the exponent term $q$ and the Yang-Mills charge parameter $Q$ are completely opposite. In the near future, with significant data from next-generation gravitational wave detectors like LISA, it might be possible to have a stringent bound on the Yang-Mills field with the help of quasinormal modes.

\section{Optical Behaviour of the Black hole} \label{sec04}
\subsection{Shadow}\label{sec04A}

This section delves into the subject of the shadow cast by black holes, which refers to the dark regions of space-time enveloping the event horizon of these astronomical phenomena. Black holes are characterized by their tremendously strong gravitational force, which is so powerful that not even light can evade its grasp once it crosses the event horizon \cite{EslamPanah:2020hoj, Konoplya2019}. Consequently, the black hole's periphery creates an obscure and gloomy circle against the backdrop of surrounding matter or light. The size and shape of this shadowy figure can provide critical information regarding the properties of black holes and the nature of gravity. With the aid of recent astronomical and technical advancements, it is now feasible to capture images of black hole shadows, which has significantly enhanced our understanding of these enigmatic entities \cite{gogoi_joulethomson_2023}.

We start with the Euler-Lagrange equation
\begin{equation}\label{shadow1}
\frac{d}{d\tau}\!\left(\frac{\partial\mathcal{L}}{\partial\dot{x}^{\mu}}\right)-\frac{\partial\mathcal{L}}{\partial x^{\mu}}=0,
\end{equation}
where the Lagrangian is of the form,
\begin{equation}\label{shadow11}
\mathcal{L}(x,\dot{x})=\frac{1}{2}\,g_{\mu\nu}\dot{x}^{\mu}\dot{x}^{\nu}.
\end{equation}
When one considers the static and spherically symmetric spacetime metric, this Lagrangian takes the form 
\begin{equation}\label{shadow2}
\mathcal{L}(x,\dot{x})=\frac{1}{2}\left[-f(r)\,\dot{t}^{2}+\frac{1}{f(r)}\,\dot{r}^{2}+r^{2}\left(\dot{\theta}^{2}+\sin^{2}\theta\dot{\phi}^{2}\right)\right].
\end{equation}
In the above expressions, the dot over the variable denotes the derivative with respect to $\tau$, the proper time.

When we opt for the equatorial plane where $\theta=\pi/2$, $\mathcal{E}$, the conserved energy and $L$, the angular momentum can be evaluated using killing vectors $\partial/\partial \tau$ and $\partial/\partial \phi$ as
\begin{equation}
\mathcal{E}=f(r)\,\dot{t},\quad L=r^{2}\dot{\phi}.
\end{equation}
For photons, the geodesic equation can be expressed as,
\begin{equation}\label{eq22}
-f(r)\,\dot{t}^{2}+\frac{\dot{r}^{2}}{f(r)}\,+r^{2}\dot{\phi}^{2} = 0.
\end{equation}
Using Eq.\ \eqref{eq22} together with the conserved quantities {\it i.e.}, \ $\mathcal{E}$ and $L$ the orbital equation of photon can be determined as \cite{gogoi_joulethomson_2023}
\begin{equation}\label{eff}
\left(\frac{dr}{d\phi}\right)^{2}=V_{eff},
\end{equation}
here the effective potential $V_{eff}$ can be defined as
\begin{equation}
V_{eff}= r^{4} \left[\frac{\mathcal{E}^{2}}{L^{2}}-\frac{f(r)}{r^{2}}\right].
\end{equation}
In radial form rewriting Eq.\ \eqref{eff} will result,
\begin{equation}
    V_r(r) = \dfrac{1}{\xi^2} - \dot{r}^2/L^2.
\end{equation}
In this expression, $\xi$ is the impact parameter and can be given as $\xi=L/\mathcal{E}$. Also the reduced potential $V_r(r)$ can be given as
\begin{equation}\label{pot}
V_r(r) = \frac{f(r)}{ r^2}.
\end{equation}

In order to examine the shadow of black holes, we direct our attention to a specific point along the trajectory, denoted as $r_{ph}$. This point corresponds to the turning point, marking the location of the light ring that encircles the black hole. It can also be regarded as the radius of the photon sphere, which is of particular interest in analyzing the characteristics of the black hole's shadow.
At this turning point of a black hole,
\cite{Jafarzade:2020ova, EslamPanah:2020hoj, pantig_shadow_2022, Ovgun:2018tua, Papnoi:2014aaa, Ovgun:2019jdo, gogoi_joulethomson_2023}
\begin{equation}
\left.V_{eff}\right|_{r_{ph}}\!\!\!\!=0,\;\; \text{and}\;\;\; 
\left.V_{eff}\right|_{r_{ph}}\!\!\!\!=0.
\end{equation} 
It is possible to obtain the impact parameter $\xi$ at the turning point as
\begin{equation}
\frac{1}{\xi_{crit}^{2}}=\frac{f(r_{ph})}{r_{ph}^{2}}.
\label{impact}
\end{equation}
Thus the radius of the photon sphere $r_{ph}$ can be found using \cite{gogoi_joulethomson_2023}: 
\begin{equation}
\left.\frac{d}{dr}\,\mathcal{A}(r)\right|_{r_{ph}}\!\!\!\!\!\!\! = 0.
\end{equation}
This equation can be rewritten as
\begin{equation}
\frac{f^{\prime}(r_{ph})}{f(r_{ph})}-\frac{h^{\prime}(r_{ph})}{h(r_{ph})}=0,
\label{photon}
\end{equation}
here $\mathcal{A}(r)=h(r)/f(r)$ with $h(r)=r^{2}$. 

Now, in order to obtain the shadow of the black hole, we 
rewrite Eq.\ \eqref{eff} with Eq.\ \eqref{impact}, in terms of the 
function $\mathcal{A}(r)$ as 
\begin{equation}
\left(\frac{dr}{d\phi}\right)^{\!2}= h(r)f(r)\left(\frac{\mathcal{A}(r)}{\mathcal{A}(r_{ph})}-1\right).
\label{eq33}
\end{equation}
Using this Eq. \eqref{eq33}, one can obtain the shadow radius of the black hole.
When examining the scenario where a static observer is located at a distance $r_0$ from the black hole, we can determine the angle $\alpha$ between the light rays emanating from the observer and the radial direction of the photon sphere. This angle can be calculated as \cite{gogoi_joulethomson_2023}
\begin{equation}
\cot\alpha=\frac{1}{\sqrt{f(r)h(r)}}\left.\frac{dr}{d\phi}\right|_{r\,=\,r_{0}}\!\!\!\!\!\!\!\!\!\!\!.
\end{equation}

Together with Eq.\ \eqref{eq33}, we can express the above equation as 
\begin{equation}
\cot^{2}\!\alpha=\frac{\mathcal{A}(r_{0})}{\mathcal{A}(r_{ph})}-1.
\end{equation}
Again, above equation can be rewritten  using the relation $\sin^{2}\!\alpha=1/(1+\cot^{2}\!\alpha)$ as
\begin{equation}
\sin^{2}\!\alpha=\frac{\mathcal{A}(r_{ph})}{\mathcal{A}(r_{0})}.
\end{equation}

Substitution of the form of $\mathcal{A}(r_{ph})$ from Eq.\ \eqref{impact} 
and $\mathcal{A}(r_{0}) = r_0^2/f(r_0)$ the black hole's shadow radius 
for a static observer at $r_{0}$ is estimated as \cite{gogoi_joulethomson_2023} 
\begin{equation}\label{shadoweq}
R_{s}=r_{0}\sin\alpha=\sqrt{\frac{r_{ph}^2f(r_{0})}{f\left(r_{ph}\right)}}.
\end{equation}
{ Again, for asymptotically flat spacetime at $r_0 \rightarrow \infty$ i.e.\ for a static observer at large distance, 
$f(r_0) \rightarrow 1$, so for such an observer the shadow radius $R_s$ becomes, 
\begin{equation}
R_{s} = \frac{r_{ph}}{\sqrt{f(r_{ph})}}.
\end{equation}

But in this scenario, the black hole spacetime is not asymptotically flat for $\Lambda \neq 0$. Hence, the black hole shadow will have an observer distance dependency. In this case, one should note that for dS spacetime, the black hole can have two horizons viz., event horizon and cosmological horizon. Any physical observer is expected to lie in between these two horizons.

Finally, via the stereographic projection of the shadow from the 
black hole's plane to the observer's image plane with coordinates 
$(X,Y)$, the apparent form of the shadow can be determined. These 
coordinates are defined as \cite{gogoi_joulethomson_2023, karmakar_thermodynamics_2023, Anacleto:2021qoe} 
\begin{align}
 X & =-\,r_{0}^{2}\sin\theta_{0}\left.\frac{d\phi}{dr}\right|_{r_{0}},\\[5pt]
Y & =r_{0}^{2}\left.\frac{d\theta}{dr}\right|_{(r_{0},\theta_{0})},
\end{align}
here $\theta_{0}$ represents the angular position of the observer with respect to the plane of the black hole. }

\begin{figure*}[t!]
      	\centering{
      	\includegraphics[scale=0.55]{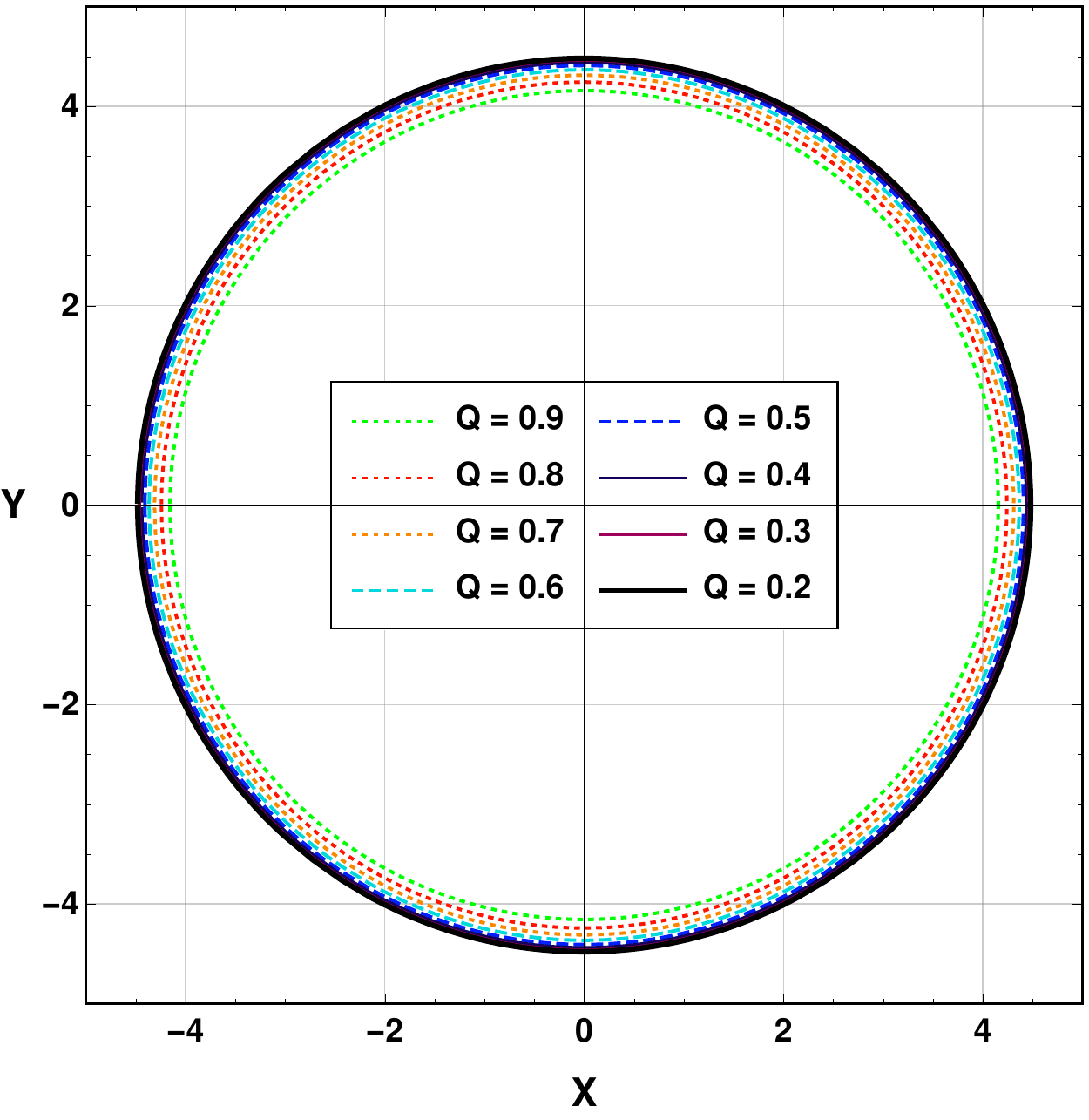} \hspace{1cm}
       \includegraphics[scale=0.55]{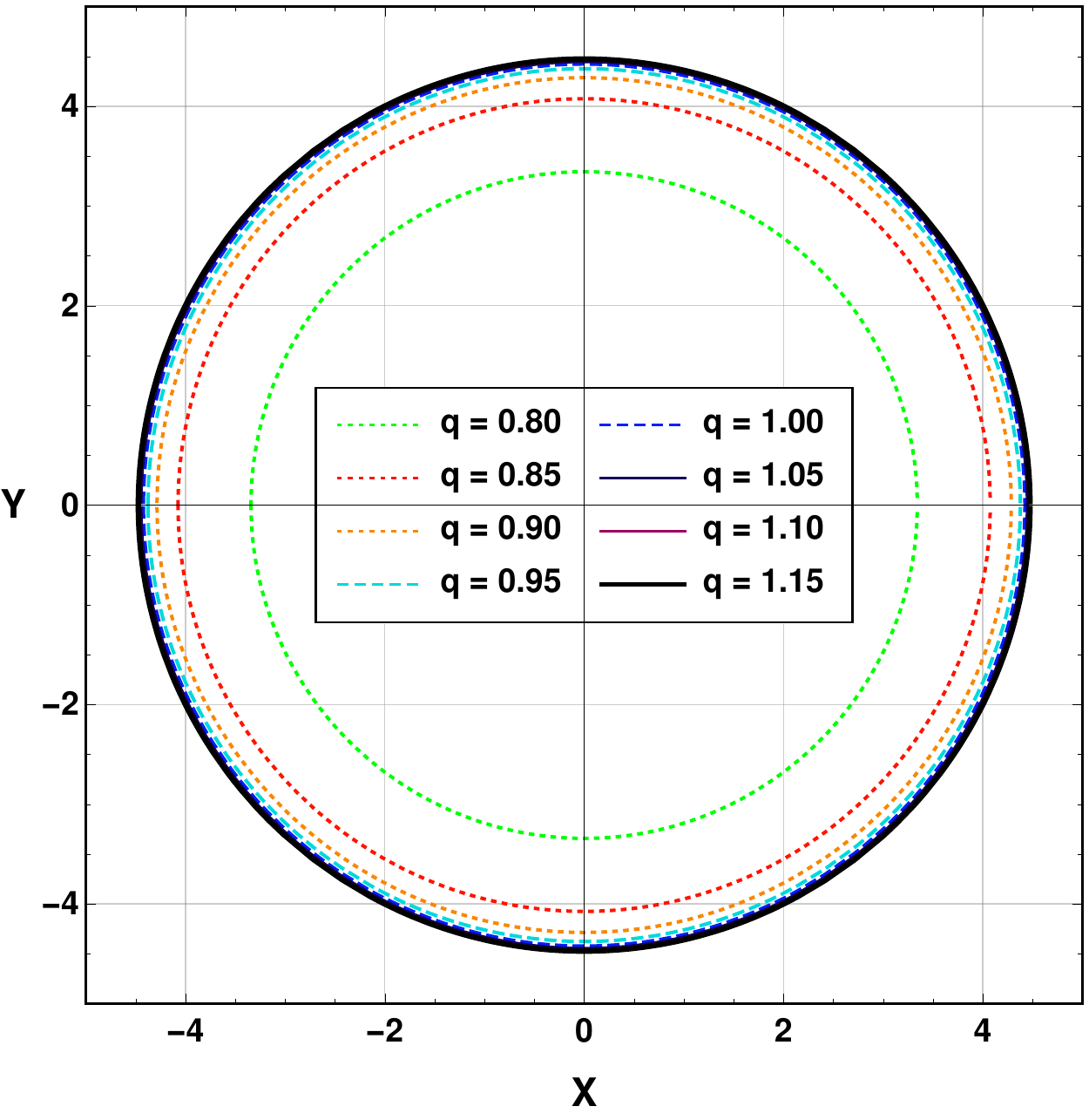}}
      	\caption{Stereographic projection of black hole shadow using observer distance $r_0 = 10,$ $M = 1$ and $\Lambda = 0.002$. On the first panel, we have used $Q = 0.3$ and on the second panel $q = 1.1$. }
      	\label{figSh01}
      \end{figure*}

\begin{figure}[t!]
      	\centering{
      	\includegraphics[scale=0.55]{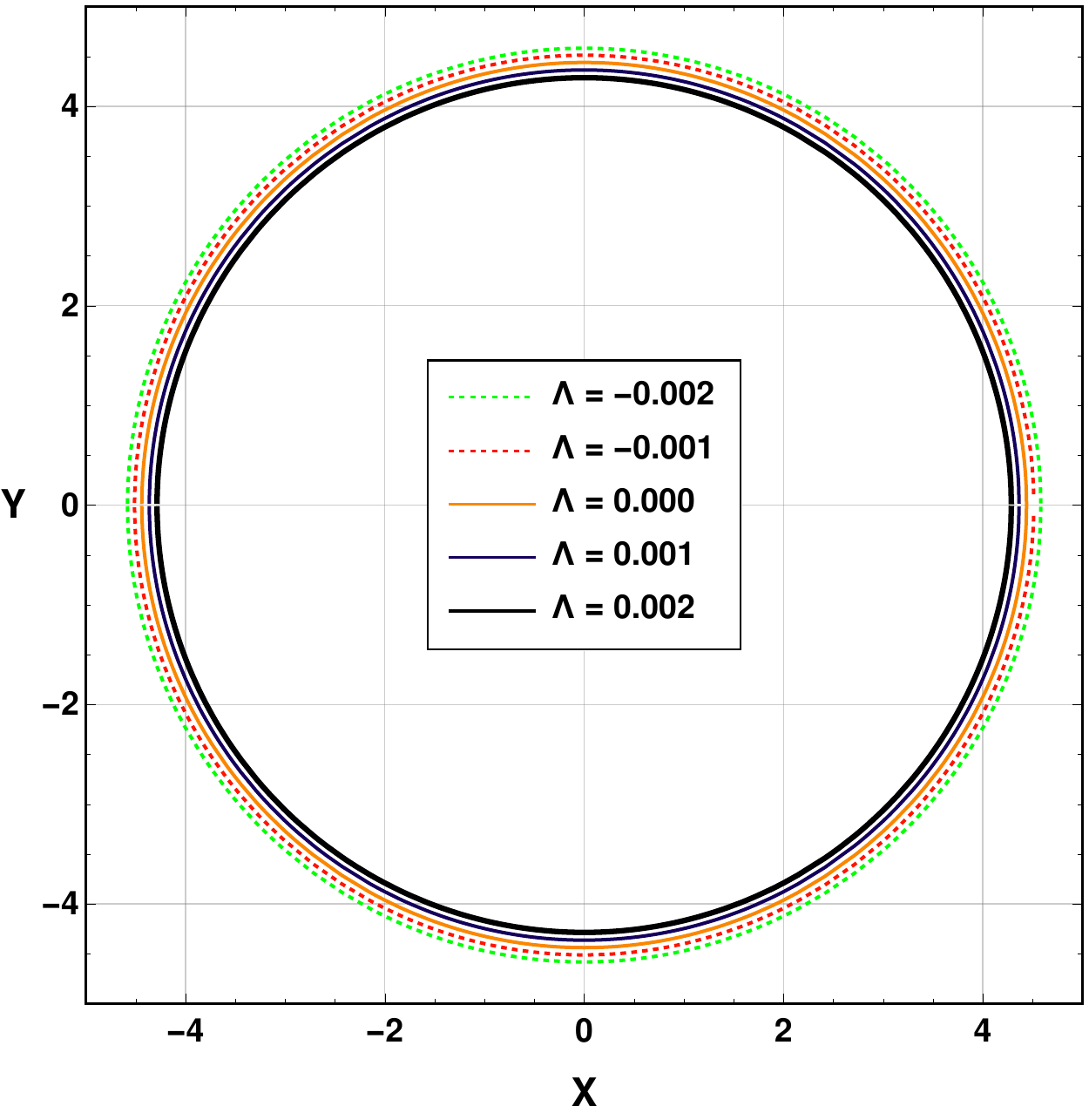}}
      	\caption{Stereographic projection of black hole shadow using observer distance $r_0 = 10,$ $M = 1$, $Q=0.3$ and $q=0.9$. }
      	\label{figSh02}
      \end{figure}

The investigation presents the stereographic projections of the black hole's shadow, offering visual representations that depict the shadow's characteristics for the specific black hole being examined. Fig. \ref{figSh01} showcases these projections, providing insights into the shadow's behaviour across various model parameters. In the initial panel of Fig. \ref{figSh01}, the stereographic projections illustrate the black hole's shadow for different values of the model parameter $Q$. Remarkably, as the model parameter $Q$ increases, a discernible reduction in the black hole's shadow is observed.

Moving to the second panel, we observe the shadow of the black hole for different values of the parameter $q$. It becomes evident that an increase in the parameter $q$ leads to a nonlinear growth in the black hole's shadow. Particularly noteworthy is the significant impact of small variations in $q$ around the value of $0.80$, resulting in a substantial alteration in the black hole's shadow. However, as $q$ approaches larger values near $1$, the effect on the black hole's shadow becomes increasingly marginal. Thus, this investigation demonstrates that both model parameters exert distinct influences on the black hole's shadow.

{ Finally, in Fig. \ref{figSh02}, we have shown the effect of the cosmological constant $\Lambda$ on the black hole shadow. One can see that for the dS black hole shadow is of smaller size while for the AdS case, the black hole shadow is comparatively larger.}

\subsection{Emission rate} \label{sec04B}

Through the examination of the black hole shadow, it is possible to investigate the emission of particles in the vicinity of the black hole. Studies have demonstrated that, for an observer positioned far away, the black hole shadow is an indication of the black hole's high-energy absorption cross-section \cite{Wei:2013kza}. Generally speaking, when dealing with a spherically symmetric black hole, the absorption cross-section shows oscillatory behaviour around a constant limiting value of $\sigma_{lim}$ at extremely high energies. Since the shadow provides a way to visually detect a black hole, it is roughly equivalent to the area of the photon sphere, which can be estimated as $\sigma_{lim} \approx \pi R_{s}^{2}$. The energy emission 
rate can be calculated using the following expression \cite{Wei:2013kza, EslamPanah:2020hoj, Kruglov:2021qzd, Kruglov:2021stm}:

\begin{equation}\label{Eqemission}
\frac{d^{2}\mathcal{E}(\omega )}{dtd\omega }=\frac{2\pi ^{3}\omega
^{3}R_{s}^{2}}{e^{\frac{\omega }{T}}-1}, 
\end{equation}%
in which $\omega $ is the emission frequency, and $T$ is the Hawking temperature given by
\begin{equation}
T= \dfrac{ f'(r_h)}{4 \pi},  \label{EqTH}
\end{equation}
where $r_h$ represents the event horizon of the black hole.

\begin{figure*}[t!]
      	\centering{
      	\includegraphics[scale=0.65]{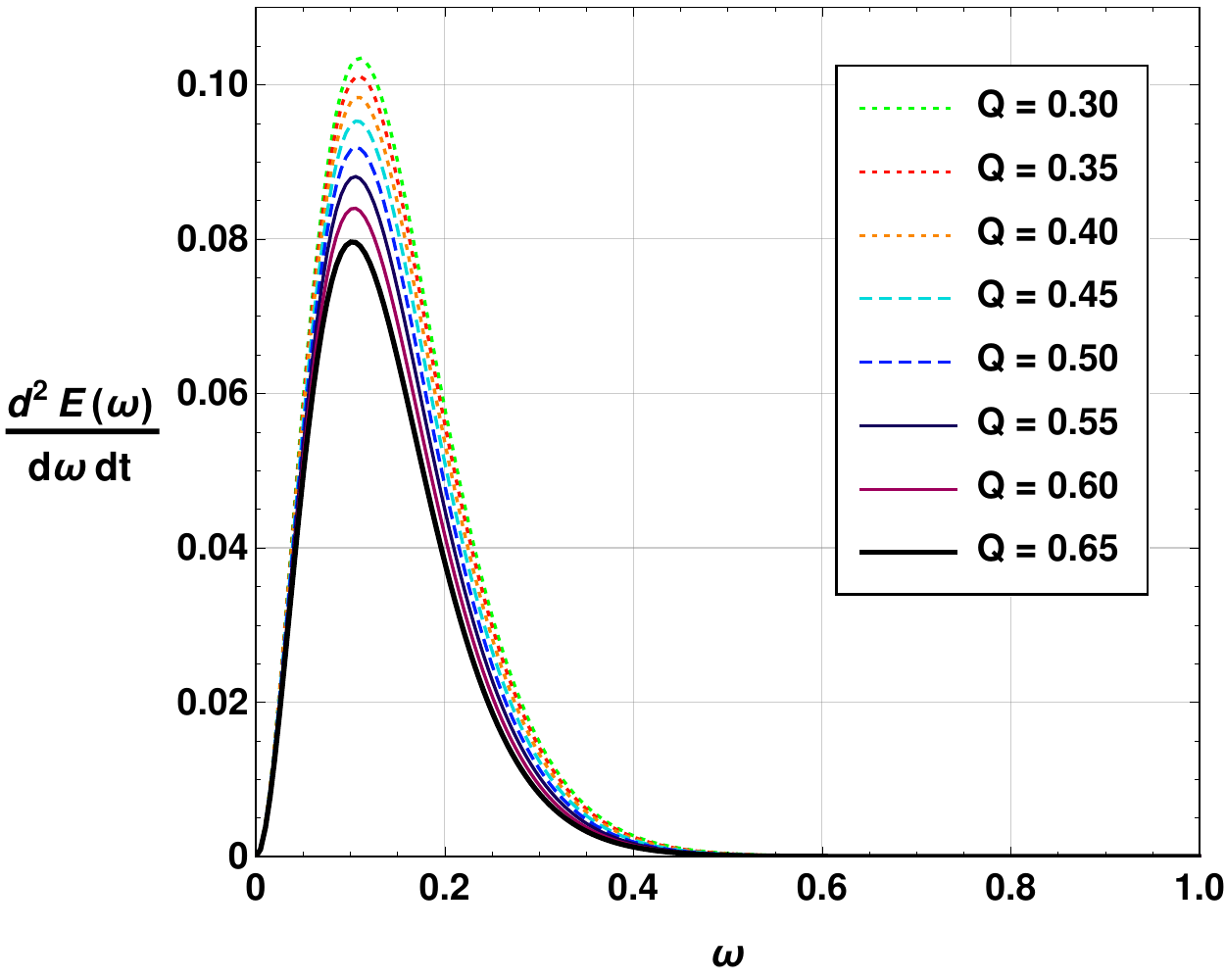}
       \includegraphics[scale=0.65]{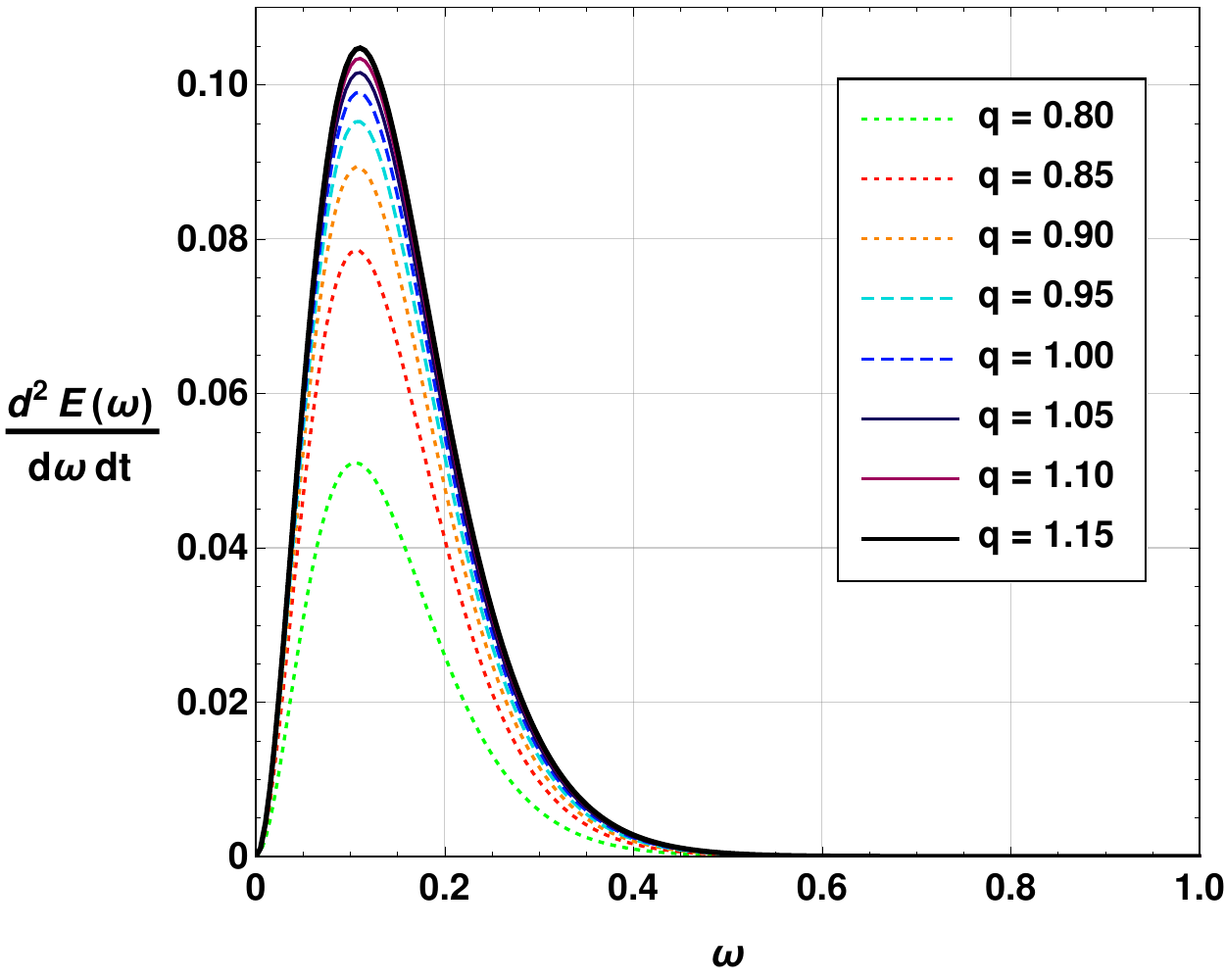}}
      	\caption{Emission rate of the black hole  with respect to $\omega$ using observer distance $r_0 = 10,$ $M = 1$ and $\Lambda = 0.002$. On the first panel, we have used $q = 1.1$ and on the second panel $Q = 0.3$.}
      	\label{figEmission01}
      \end{figure*}

\begin{figure}[htb!]
      	\centering{
      	\includegraphics[scale=0.65]{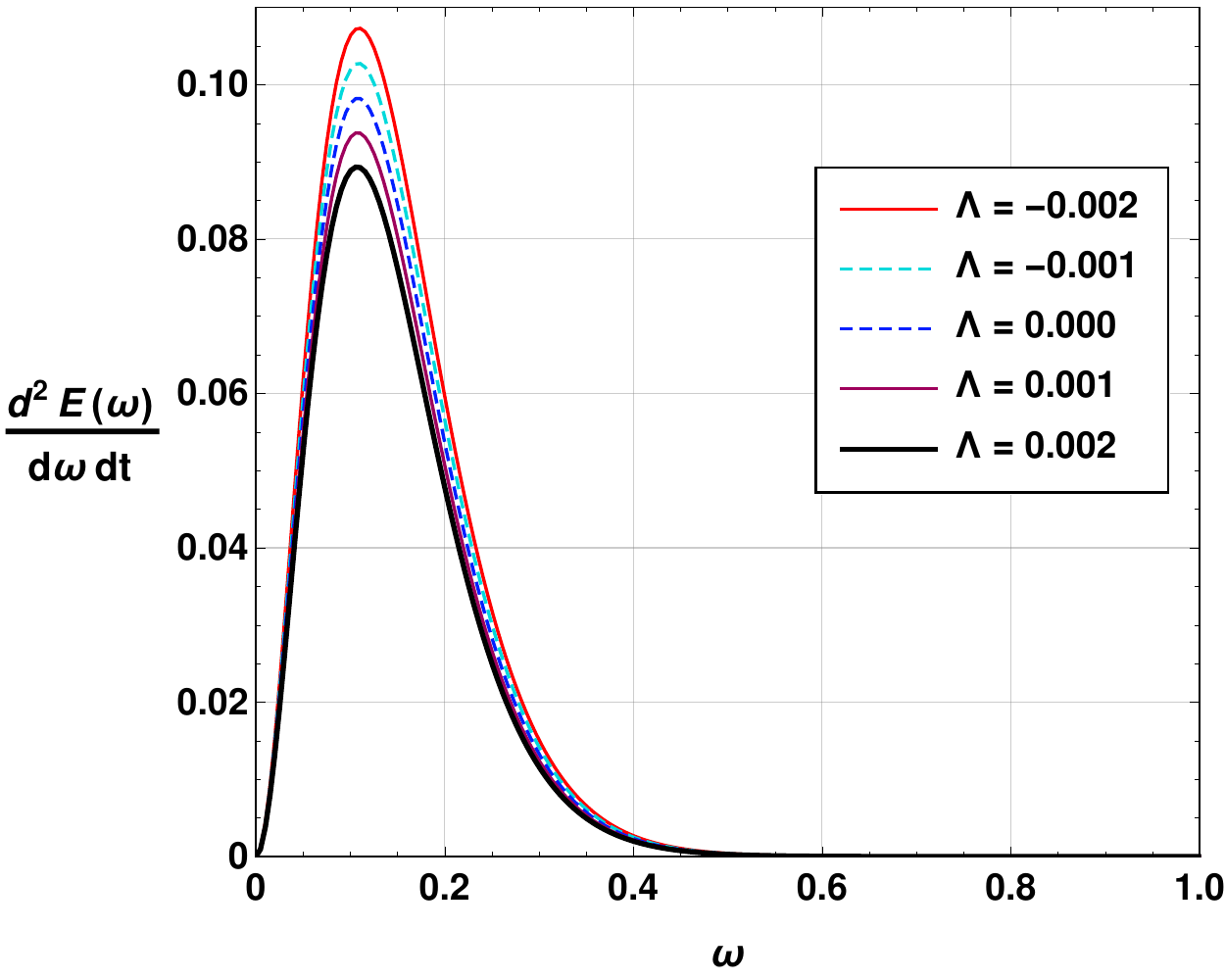}}
      	\caption{Emission rate of the black hole  with respect to $\omega$ using observer distance $r_0 = 10,$ $M = 1$, $Q=0.3$ and $q=0.9$. }
      	\label{figEmission02}
      \end{figure}

By utilizing the Hawking temperature expression provided in Eq. (\ref{Eqemission}), we have illustrated the change in emission rate concerning $\omega$ for various values of $Q$ (displayed on the first panel) and $q$ (shown on the second panel) in Fig. \ref{figEmission01}. It is noticeable that as the parameter $Q$ increases, the emission rate of the black hole reduces. In contrast, with an increase in the parameter $q$, the emission rate of the black hole increases slowly. Therefore, the examination indicates that as the Yang-Mills charge parameter $Q$ grows, the black hole's evaporation rate decreases, and its lifespan extends. Conversely, as the parameter $q$ increases, the black hole's evaporation rate gradually increases, and its lifespan shortens.

{ 
In Fig. \ref{figEmission02}, we have shown the impacts on the emission rate by the cosmological constant $\Lambda$. One can see that for the AdS case, emission rate is higher resulting short lifespan of the black hole while in the case of dS scenario, the emission rate decreases resulting a black hole with a longer lifespan.
}

\section{Connection between quasinormal modes and optical properties of the black hole}
\label{sec05}
{ 
The shadow of the black hole is associated with the quasinormal modes. An investigation of their connection can be helpful to make predictions of the quasinormal mode behaviours from the observational results of black hole shadow. Moreover, it also can help us to put a constraint on the quasinormal modes from the black hole shadow observations. To obtain an approximate relation of the black hole shadow with quasinormal modes, we can use the WKB formula. However, higher-order WKB formula can result in a more complicated relation. In this study, we are primarily interested to see how they are related in the eikonal regime. Hence, although we have used 6th-order WKB approximations in the previous part of the investigation, here we can simply use 3rd-order WKB approximation, as both formulas will give us the same leading order approximations in the eikonal regime.

}

At first, we start with the expression of the third-order WKB expansion of quasinormal mode frequency as follows \cite{Konoplya:2011qq, Cuadros-Melgar:2020kqn}:
\begin{widetext}
\begin{eqnarray}
\nonumber
\omega = \Bigg\{  V + \frac{V_4}{8V_2}\left( \nu^2+\frac{1}{4}\right) - \left( \frac{7 + 60\nu^2}{288}\right) \frac{V_3^2}{V_2^2}+i\nu \sqrt{-2V_2}\left[\frac{1}{2V_2}\left[\frac{5V_3^4(77+188\nu^2)}{6912V_2^4} \right. \right. \\
\nonumber
\left. \left. -\frac{V_3^2V_4(51+100\nu^2 )}{384V_2^3}+ \frac{V_4^2(67+68\nu^2 )}{2304V_2^2}+\frac{V_5V_3(19+28\nu^2 )}{288V_2^2}+\frac{V_6(5+4\nu^2 )}{288V_2}\right] -1 \right]  \Bigg\}^{1/2}_{r=r_{ph}}.\\
\label{e7}
\end{eqnarray}
\end{widetext}
In this context, $V_i$ denotes the derivative of the potential $V$ with respect to its $i$-th variable. Furthermore, $\nu$ is defined as the overtone number, represented by $n+1/2$, where $n$ is a natural number.
The expression for the frequency $\omega$ mentioned above should be evaluated at the specific point $r_{ph}$, which is defined as the maximum value attained by the potential $V$ throughout the system. { One may note that this relation can provide the quasinormal modes up to a satisfactory accuracy. However, this formula can result in larger errors when the overtone number $n$ is greater than the multipole number $l$. The error decreases significantly when $l>>n$.}

Eventually, when we expand Eq. (\ref{e7}), we acquire, specifically in the eikonal regime \cite{Cuadros-Melgar:2020kqn},
\begin{widetext}
\begin{eqnarray}
\nonumber
\omega =  \left[(2\ell+1) \frac{\sqrt{f(r)}}{2r} + \mathcal{O}(\ell^{-1})\right]_{r=r_{ph}} -
 i \left[\frac{n+1/2}{\sqrt{2}}\frac{\sqrt{f(r)}}{r} 
\sqrt{6rf'(r)-6f(r)-r^2f''(r)-r^2f(r)^{-1}{f'(r)}^2} +  \mathcal{O}(\ell^{-1})\right]_{r=r_{ph}}.
\label{e10}
\end{eqnarray} 
\end{widetext}
The leading-order approximation of the imaginary part can be expressed as follows \cite{Cuadros-Melgar:2020kqn}:
\begin{equation}
\label{imgR}
 \omega_I = \frac{n +1}{\sqrt{2}}R_s^{-1}\sqrt{2f(r_{ph})-r_{ph}^2f''(r_{ph})} +\mathcal{O}(\ell^{-1}).
\end{equation}

Regarding the real part, in the leading order approximation, it precisely corresponds to the shadow radius of the black hole. However, in the sub-leading regime, it is equal to half of the shadow radius's value. The leading-order approximations provide \cite{Cuadros-Melgar:2020kqn}
\begin{equation}
\label{realR}
\omega_R = R_s^{-1}\left(\ell + \frac{1}{2} +\mathcal{O}(\ell^{-1})\right).
\end{equation}

From Eq.s \ref{imgR} and \ref{realR}, one can see that at the eikonal regime, the shadow radius of the black hole is inversely connected with the damping rate and GW oscillation frequency. Hence, at the eikonal limit, it is possible to constrain the damping rate and the GW oscillation frequency with the help of black hole shadow. We leave it as a future prospect of this work. Moreover, Eq. \eqref{Eqemission} shows that the emission rate is associated with the square of the black hole shadow radius, hence, knowing the Hawking temperature, it is possible to comment on the emission rate from the quasinormal modes. In the near future, from significant observational results from LISA, one might be able to have a strong constraint on the quasinormal modes which will also shed some light on the emission rate of the black hole as well as the stability.

\section{Concluding Remarks}\label{sec06}

Our investigation has unveiled that the Yang-Mills charge parameter, denoted as $Q$, and the exponent term, denoted as $q$, exert substantial influences on the quasinormal spectrum and optical characteristics of a four-dimensional black hole governed by Einstein Power-Yang-Mills gravity. Remarkably, these two parameters exhibit diametrically opposite effects on the quasinormal mode spectrum.

To gain a clearer understanding of the behaviour of these Yang-Mills field parameters, we have examined the black hole's shadow and emission rate. Our findings indicate that the black hole's shadow size decreases as the Yang-Mills charge parameter $Q$ increases. Conversely, the exponent parameter $q$ shows contrasting behaviour, as an increase in its value leads to an augmented shadow. This variation is particularly prominent when $q$ approaches its lower boundary of 0.75. Additionally, we have analyzed the black hole's emission rate. As the Yang-Mills charge parameter $Q$ rises, the black hole's evaporation rate decreases. { On the other hand, an increase in the value of $q$ increases the evaporation rate, indicating a reduced lifespan for the black hole.}

Therefore, our research underscores the significant impact of both the Yang-Mills charge parameter and the exponent term on the quasinormal spectrum and optical properties of a four-dimensional black hole governed by Einstein Power-Yang-Mills gravity.

\section*{Acknowledgments}
DJG acknowledges the contribution of the COST Action CA21136  -- ``Addressing observational tensions in cosmology with systematics and fundamental physics (CosmoVerse)". DJG and JB acknowledge Prof. U. D. Goswami, Department of Physics, Dibrugarh University for some useful discussions.



\begin{thebibliography}{91}%
\makeatletter
\providecommand \@ifxundefined [1]{%
 \@ifx{#1\undefined}
}%
\providecommand \@ifnum [1]{%
 \ifnum #1\expandafter \@firstoftwo
 \else \expandafter \@secondoftwo
 \fi
}%
\providecommand \@ifx [1]{%
 \ifx #1\expandafter \@firstoftwo
 \else \expandafter \@secondoftwo
 \fi
}%
\providecommand \natexlab [1]{#1}%
\providecommand \enquote  [1]{``#1''}%
\providecommand \bibnamefont  [1]{#1}%
\providecommand \bibfnamefont [1]{#1}%
\providecommand \citenamefont [1]{#1}%
\providecommand \href@noop [0]{\@secondoftwo}%
\providecommand \href [0]{\begingroup \@sanitize@url \@href}%
\providecommand \@href[1]{\@@startlink{#1}\@@href}%
\providecommand \@@href[1]{\endgroup#1\@@endlink}%
\providecommand \@sanitize@url [0]{\catcode `\\12\catcode `\$12\catcode
  `\&12\catcode `\#12\catcode `\^12\catcode `\_12\catcode `\%12\relax}%
\providecommand \@@startlink[1]{}%
\providecommand \@@endlink[0]{}%
\providecommand \url  [0]{\begingroup\@sanitize@url \@url }%
\providecommand \@url [1]{\endgroup\@href {#1}{\urlprefix }}%
\providecommand \urlprefix  [0]{URL }%
\providecommand \Eprint [0]{\href }%
\providecommand \doibase [0]{http://dx.doi.org/}%
\providecommand \selectlanguage [0]{\@gobble}%
\providecommand \bibinfo  [0]{\@secondoftwo}%
\providecommand \bibfield  [0]{\@secondoftwo}%
\providecommand \translation [1]{[#1]}%
\providecommand \BibitemOpen [0]{}%
\providecommand \bibitemStop [0]{}%
\providecommand \bibitemNoStop [0]{.\EOS\space}%
\providecommand \EOS [0]{\spacefactor3000\relax}%
\providecommand \BibitemShut  [1]{\csname bibitem#1\endcsname}%
\let\auto@bib@innerbib\@empty
\bibitem [{\citenamefont {Abbott~et al.}(2016)}]{PhysRevLett.116.061102}%
  \BibitemOpen
  \bibfield  {author} {\bibinfo {author} {\bibfnamefont {B.~P.}\ \bibnamefont
  {Abbott~et al.}} (\bibinfo {collaboration} {LIGO Scientific Collaboration and
  Virgo Collaboration}),\ }\bibfield  {title} {\enquote {\bibinfo {title}
  {Observation of gravitational waves from a binary black hole merger},}\
  }\href {\doibase 10.1103/PhysRevLett.116.061102} {\bibfield  {journal}
  {\bibinfo  {journal} {Phys. Rev. Lett.}\ }\textbf {\bibinfo {volume} {116}},\
  \bibinfo {pages} {061102} (\bibinfo {year} {2016})}\BibitemShut {NoStop}%
\bibitem [{\citenamefont {Vishveshwara}(1970)}]{Vishveshwara:1970zz}%
  \BibitemOpen
  \bibfield  {author} {\bibinfo {author} {\bibfnamefont {C.~V.}\ \bibnamefont
  {Vishveshwara}},\ }\bibfield  {title} {\enquote {\bibinfo {title}
  {{Scattering of Gravitational Radiation by a Schwarzschild Black-hole}},}\
  }\href {\doibase 10.1038/227936a0} {\bibfield  {journal} {\bibinfo  {journal}
  {Nature}\ }\textbf {\bibinfo {volume} {227}},\ \bibinfo {pages} {936}
  (\bibinfo {year} {1970})}\BibitemShut {NoStop}%
\bibitem [{\citenamefont {Press}(1971)}]{Press:1971wr}%
  \BibitemOpen
  \bibfield  {author} {\bibinfo {author} {\bibfnamefont {W.~H.}\ \bibnamefont
  {Press}},\ }\bibfield  {title} {\enquote {\bibinfo {title} {{Long Wave Trains
  of Gravitational Waves from a Vibrating Black Hole}},}\ }\href {\doibase
  10.1086/180849} {\bibfield  {journal} {\bibinfo  {journal} {Astrophys. J.
  Lett.}\ }\textbf {\bibinfo {volume} {170}},\ \bibinfo {pages} {L105}
  (\bibinfo {year} {1971})}\BibitemShut {NoStop}%
\bibitem [{\citenamefont {Kokkotas}\ and\ \citenamefont
  {Schmidt}(1999)}]{Kokkotas:1999bd}%
  \BibitemOpen
  \bibfield  {author} {\bibinfo {author} {\bibfnamefont {K.~D.}\ \bibnamefont
  {Kokkotas}}\ and\ \bibinfo {author} {\bibfnamefont {B.~G.}\ \bibnamefont
  {Schmidt}},\ }\bibfield  {title} {\enquote {\bibinfo {title} {{Quasinormal
  modes of stars and black holes}},}\ }\href {\doibase 10.12942/lrr-1999-2}
  {\bibfield  {journal} {\bibinfo  {journal} {Living Rev. Rel.}\ }\textbf
  {\bibinfo {volume} {2}},\ \bibinfo {pages} {2} (\bibinfo {year} {1999})},\
  \Eprint {http://arxiv.org/abs/gr-qc/9909058}{arXiv:gr-qc/9909058}\BibitemShut
  {NoStop}%
\bibitem [{\citenamefont {Rinc\'on}\ and\ \citenamefont
  {Panotopoulos}(2018)}]{Rincon:2018ktz}%
  \BibitemOpen
  \bibfield  {author} {\bibinfo {author} {\bibfnamefont {A.}~\bibnamefont
  {Rinc\'on}}\ and\ \bibinfo {author} {\bibfnamefont {G.}~\bibnamefont
  {Panotopoulos}},\ }\bibfield  {title} {\enquote {\bibinfo {title} {{Greybody
  factors and quasinormal modes for a nonminimally coupled scalar field in a
  cloud of strings in (2+1)-dimensional background}},}\ }\href {\doibase
  10.1140/epjc/s10052-018-6352-5} {\bibfield  {journal} {\bibinfo  {journal}
  {Eur. Phys. J. C}\ }\textbf {\bibinfo {volume} {78}},\ \bibinfo {pages} {858}
  (\bibinfo {year} {2018})},\ \Eprint
  {http://arxiv.org/abs/1810.08822}{arXiv:1810.08822 [gr-qc]}\BibitemShut
  {NoStop}%
\bibitem [{\citenamefont {Liu}\ \emph {et~al.}(2022)\citenamefont {Liu},
  \citenamefont {Yang}, \citenamefont {\"Ovg\"un}, \citenamefont {Long},\ and\
  \citenamefont {Xu}}]{Liu:2022ygf}%
  \BibitemOpen
  \bibfield  {author} {\bibinfo {author} {\bibfnamefont {D.}~\bibnamefont
  {Liu}}, \bibinfo {author} {\bibfnamefont {Y.}~\bibnamefont {Yang}}, \bibinfo
  {author} {\bibfnamefont {A.}~\bibnamefont {\"Ovg\"un}}, \bibinfo {author}
  {\bibfnamefont {Z.-W.}\ \bibnamefont {Long}}, \ and\ \bibinfo {author}
  {\bibfnamefont {Z.}~\bibnamefont {Xu}},\ }\bibfield  {title} {\enquote
  {\bibinfo {title} {{Quasinormal Modes and Greybody Bounds of Rotating Black
  Holes in a Dark Matter Halo}},}\ }\href@noop {} {\  (\bibinfo {year}
  {2022})},\ \Eprint {http://arxiv.org/abs/2204.11563}{arXiv:2204.11563
  [gr-qc]}\BibitemShut {NoStop}%
\bibitem [{\citenamefont {Rincon}\ \emph {et~al.}(2022)\citenamefont {Rincon},
  \citenamefont {Gonzalez}, \citenamefont {Panotopoulos}, \citenamefont
  {Saavedra},\ and\ \citenamefont {Vasquez}}]{Rincon:2021gwd}%
  \BibitemOpen
  \bibfield  {author} {\bibinfo {author} {\bibfnamefont {A.}~\bibnamefont
  {Rincon}}, \bibinfo {author} {\bibfnamefont {P.~A.}\ \bibnamefont
  {Gonzalez}}, \bibinfo {author} {\bibfnamefont {G.}~\bibnamefont
  {Panotopoulos}}, \bibinfo {author} {\bibfnamefont {J.}~\bibnamefont
  {Saavedra}}, \ and\ \bibinfo {author} {\bibfnamefont {Y.}~\bibnamefont
  {Vasquez}},\ }\bibfield  {title} {\enquote {\bibinfo {title} {{Quasinormal
  modes for a non-minimally coupled scalar field in a five-dimensional
  Einstein\textendash{}Power\textendash{}Maxwell background}},}\ }\href
  {\doibase 10.1140/epjp/s13360-022-03438-4} {\bibfield  {journal} {\bibinfo
  {journal} {Eur. Phys. J. Plus}\ }\textbf {\bibinfo {volume} {137}},\ \bibinfo
  {pages} {1278} (\bibinfo {year} {2022})},\ \Eprint
  {http://arxiv.org/abs/2112.04793}{arXiv:2112.04793 [gr-qc]}\BibitemShut
  {NoStop}%
\bibitem [{\citenamefont {Ovg\"un}\ and\ \citenamefont
  {Jusufi}(2018)}]{Ovgun:2018gwt}%
  \BibitemOpen
  \bibfield  {author} {\bibinfo {author} {\bibfnamefont {A.}~\bibnamefont
  {Ovg\"un}}\ and\ \bibinfo {author} {\bibfnamefont {K.}~\bibnamefont
  {Jusufi}},\ }\bibfield  {title} {\enquote {\bibinfo {title} {{Quasinormal
  Modes and Greybody Factors of $f(R)$ gravity minimally coupled to a cloud of
  strings in $2+1$ Dimensions}},}\ }\href {\doibase 10.1016/j.aop.2018.05.013}
  {\bibfield  {journal} {\bibinfo  {journal} {Annals Phys.}\ }\textbf {\bibinfo
  {volume} {395}},\ \bibinfo {pages} {138} (\bibinfo {year} {2018})},\ \Eprint
  {http://arxiv.org/abs/1801.02555}{arXiv:1801.02555 [gr-qc]}\BibitemShut
  {NoStop}%
\bibitem [{\citenamefont {\"Ovg\"un}\ \emph {et~al.}(2021)\citenamefont
  {\"Ovg\"un}, \citenamefont {Sakall\i{}},\ and\ \citenamefont
  {Mutuk}}]{Ovgun:2019yor}%
  \BibitemOpen
  \bibfield  {author} {\bibinfo {author} {\bibfnamefont {A.}~\bibnamefont
  {\"Ovg\"un}}, \bibinfo {author} {\bibfnamefont {I.}~\bibnamefont
  {Sakall\i{}}}, \ and\ \bibinfo {author} {\bibfnamefont {H.}~\bibnamefont
  {Mutuk}},\ }\bibfield  {title} {\enquote {\bibinfo {title} {{Quasinormal
  modes of dS and AdS black holes: Feedforward neural network method}},}\
  }\href {\doibase 10.1142/S0219887821501541} {\bibfield  {journal} {\bibinfo
  {journal} {Int. J. Geom. Meth. Mod. Phys.}\ }\textbf {\bibinfo {volume}
  {18}},\ \bibinfo {pages} {2150154} (\bibinfo {year} {2021})},\ \Eprint
  {http://arxiv.org/abs/1904.09509}{arXiv:1904.09509 [gr-qc]}\BibitemShut
  {NoStop}%
\bibitem [{\citenamefont {Anacleto}\ \emph {et~al.}(2021)\citenamefont
  {Anacleto}, \citenamefont {Campos}, \citenamefont {Brito},\ and\
  \citenamefont {Passos}}]{Anacleto:2021qoe}%
  \BibitemOpen
  \bibfield  {author} {\bibinfo {author} {\bibfnamefont {M.~A.}\ \bibnamefont
  {Anacleto}}, \bibinfo {author} {\bibfnamefont {J.~A.~V.}\ \bibnamefont
  {Campos}}, \bibinfo {author} {\bibfnamefont {F.~A.}\ \bibnamefont {Brito}}, \
  and\ \bibinfo {author} {\bibfnamefont {E.}~\bibnamefont {Passos}},\
  }\bibfield  {title} {\enquote {\bibinfo {title} {{Quasinormal modes and
  shadow of a Schwarzschild black hole with GUP}},}\ }\href {\doibase
  10.1016/j.aop.2021.168662} {\bibfield  {journal} {\bibinfo  {journal} {Annals
  Phys.}\ }\textbf {\bibinfo {volume} {434}},\ \bibinfo {pages} {168662}
  (\bibinfo {year} {2021})},\ \Eprint
  {http://arxiv.org/abs/2108.04998}{arXiv:2108.04998 [gr-qc]}\BibitemShut
  {NoStop}%
\bibitem [{\citenamefont {Lambiase}\ \emph {et~al.}(2023)\citenamefont
  {Lambiase}, \citenamefont {Pantig}, \citenamefont {Gogoi},\ and\
  \citenamefont {\"Ovg\"un}}]{Lambiase:2023hng}%
  \BibitemOpen
  \bibfield  {author} {\bibinfo {author} {\bibfnamefont {G.}~\bibnamefont
  {Lambiase}}, \bibinfo {author} {\bibfnamefont {R.~C.}\ \bibnamefont
  {Pantig}}, \bibinfo {author} {\bibfnamefont {D.~J.}\ \bibnamefont {Gogoi}}, \
  and\ \bibinfo {author} {\bibfnamefont {A.}~\bibnamefont {\"Ovg\"un}},\
  }\bibfield  {title} {\enquote {\bibinfo {title} {{Investigating the
  Connection between Generalized Uncertainty Principle and Asymptotically Safe
  Gravity in Black Hole Signatures through Shadow and Quasinormal Modes}},}\
  }\href@noop {} {\  (\bibinfo {year} {2023})},\ \Eprint
  {http://arxiv.org/abs/2304.00183}{arXiv:2304.00183 [gr-qc]}\BibitemShut
  {NoStop}%
\bibitem [{\citenamefont {Sekhmani}\ and\ \citenamefont
  {Gogoi}(2023)}]{sekhmani_electromagnetic_2023}%
  \BibitemOpen
  \bibfield  {author} {\bibinfo {author} {\bibfnamefont {Y.}~\bibnamefont
  {Sekhmani}}\ and\ \bibinfo {author} {\bibfnamefont {D.~J.}\ \bibnamefont
  {Gogoi}},\ }\bibfield  {title} {\enquote {\bibinfo {title} {Electromagnetic
  quasinormal modes of dyonic {AdS} black holes with quasitopological
  electromagnetism in a {Horndeski} gravity theory mimicking {EGB} gravity at
  {D} → 4},}\ }\href {\doibase 10.1142/S0219887823501608} {\bibfield
  {journal} {\bibinfo  {journal} {International Journal of Geometric Methods in
  Modern Physics}\ ,\ \bibinfo {pages} {2350160}} (\bibinfo {year}
  {2023})}\BibitemShut {NoStop}%
\bibitem [{\citenamefont {Gogoi}\ \emph
  {et~al.}(2023{\natexlab{a}})\citenamefont {Gogoi}, \citenamefont
  {\"Ovg\"un},\ and\ \citenamefont {Koussour}}]{Gogoi:2023kjt}%
  \BibitemOpen
  \bibfield  {author} {\bibinfo {author} {\bibfnamefont {D.~J.}\ \bibnamefont
  {Gogoi}}, \bibinfo {author} {\bibfnamefont {A.}~\bibnamefont {\"Ovg\"un}}, \
  and\ \bibinfo {author} {\bibfnamefont {M.}~\bibnamefont {Koussour}},\
  }\bibfield  {title} {\enquote {\bibinfo {title} {{Quasinormal Modes of Black
  holes in $f(Q)$ gravity}},}\ }\href@noop {} {\  (\bibinfo {year}
  {2023}{\natexlab{a}})},\ \Eprint
  {http://arxiv.org/abs/2303.07424}{arXiv:2303.07424 [gr-qc]}\BibitemShut
  {NoStop}%
\bibitem [{\citenamefont {Parbin}\ \emph {et~al.}(2022)\citenamefont {Parbin},
  \citenamefont {Gogoi}, \citenamefont {Bora},\ and\ \citenamefont
  {Goswami}}]{Parbin:2022iwt}%
  \BibitemOpen
  \bibfield  {author} {\bibinfo {author} {\bibfnamefont {N.}~\bibnamefont
  {Parbin}}, \bibinfo {author} {\bibfnamefont {D.~J.}\ \bibnamefont {Gogoi}},
  \bibinfo {author} {\bibfnamefont {J.}~\bibnamefont {Bora}}, \ and\ \bibinfo
  {author} {\bibfnamefont {U.~D.}\ \bibnamefont {Goswami}},\ }\bibfield
  {title} {\enquote {\bibinfo {title} {{Deflection angle and quasinormal modes
  of a de Sitter black hole in $f(\mathcal{T}, \mathcal{B})$ gravity}},}\
  }\href@noop {} {\  (\bibinfo {year} {2022})},\ \Eprint
  {http://arxiv.org/abs/2211.02414}{arXiv:2211.02414 [gr-qc]}\BibitemShut
  {NoStop}%
\bibitem [{\citenamefont {Karmakar}\ \emph {et~al.}(2022)\citenamefont
  {Karmakar}, \citenamefont {Gogoi},\ and\ \citenamefont
  {Goswami}}]{karmakar_quasinormal_2022}%
  \BibitemOpen
  \bibfield  {author} {\bibinfo {author} {\bibfnamefont {R.}~\bibnamefont
  {Karmakar}}, \bibinfo {author} {\bibfnamefont {D.~J.}\ \bibnamefont {Gogoi}},
  \ and\ \bibinfo {author} {\bibfnamefont {U.~D.}\ \bibnamefont {Goswami}},\
  }\bibfield  {title} {\enquote {\bibinfo {title} {Quasinormal modes and
  thermodynamic properties of {GUP}-corrected {Schwarzschild} black hole
  surrounded by quintessence},}\ }\href {\doibase 10.1142/S0217751X22501809}
  {\bibfield  {journal} {\bibinfo  {journal} {International Journal of Modern
  Physics A}\ }\textbf {\bibinfo {volume} {37}},\ \bibinfo {pages} {2250180}
  (\bibinfo {year} {2022})}\BibitemShut {NoStop}%
\bibitem [{\citenamefont {Gogoi}\ and\ \citenamefont
  {Goswami}(2022)}]{Gogoi:2022wyv}%
  \BibitemOpen
  \bibfield  {author} {\bibinfo {author} {\bibfnamefont {D.~J.}\ \bibnamefont
  {Gogoi}}\ and\ \bibinfo {author} {\bibfnamefont {U.~D.}\ \bibnamefont
  {Goswami}},\ }\bibfield  {title} {\enquote {\bibinfo {title} {{Quasinormal
  modes and Hawking radiation sparsity of GUP corrected black holes in
  bumblebee gravity with topological defects}},}\ }\href {\doibase
  10.1088/1475-7516/2022/06/029} {\bibfield  {journal} {\bibinfo  {journal}
  {JCAP}\ }\textbf {\bibinfo {volume} {06}},\ \bibinfo {pages} {029} (\bibinfo
  {year} {2022})},\ \Eprint {http://arxiv.org/abs/2203.07594}{arXiv:2203.07594
  [gr-qc]}\BibitemShut {NoStop}%
\bibitem [{\citenamefont {Gogoi}\ \emph
  {et~al.}(2023{\natexlab{b}})\citenamefont {Gogoi}, \citenamefont {Karmakar},\
  and\ \citenamefont {Goswami}}]{Gogoi:2021cbp}%
  \BibitemOpen
  \bibfield  {author} {\bibinfo {author} {\bibfnamefont {D.~J.}\ \bibnamefont
  {Gogoi}}, \bibinfo {author} {\bibfnamefont {R.}~\bibnamefont {Karmakar}}, \
  and\ \bibinfo {author} {\bibfnamefont {U.~D.}\ \bibnamefont {Goswami}},\
  }\bibfield  {title} {\enquote {\bibinfo {title} {{Quasinormal modes of
  nonlinearly charged black holes surrounded by a cloud of strings in Rastall
  gravity}},}\ }\href {\doibase 10.1142/S021988782350007X} {\bibfield
  {journal} {\bibinfo  {journal} {Int. J. Geom. Meth. Mod. Phys.}\ }\textbf
  {\bibinfo {volume} {20}},\ \bibinfo {pages} {2350007} (\bibinfo {year}
  {2023}{\natexlab{b}})},\ \Eprint
  {http://arxiv.org/abs/2111.00854}{arXiv:2111.00854 [gr-qc]}\BibitemShut
  {NoStop}%
\bibitem [{\citenamefont {Gogoi}\ and\ \citenamefont
  {Goswami}(2021)}]{Gogoi:2021dkr}%
  \BibitemOpen
  \bibfield  {author} {\bibinfo {author} {\bibfnamefont {D.~J.}\ \bibnamefont
  {Gogoi}}\ and\ \bibinfo {author} {\bibfnamefont {U.~D.}\ \bibnamefont
  {Goswami}},\ }\bibfield  {title} {\enquote {\bibinfo {title} {{Quasinormal
  modes of black holes with non-linear-electrodynamic sources in Rastall
  gravity}},}\ }\href {\doibase 10.1016/j.dark.2021.100860} {\bibfield
  {journal} {\bibinfo  {journal} {Phys. Dark Univ.}\ }\textbf {\bibinfo
  {volume} {33}},\ \bibinfo {pages} {100860} (\bibinfo {year} {2021})},\
  \Eprint {http://arxiv.org/abs/2104.13115}{arXiv:2104.13115
  [gr-qc]}\BibitemShut {NoStop}%
\bibitem [{\citenamefont {Pantig}\ \emph
  {et~al.}(2022{\natexlab{a}})\citenamefont {Pantig}, \citenamefont
  {Mastrototaro}, \citenamefont {Lambiase},\ and\ \citenamefont
  {\"Ovg\"un}}]{Pantig:2022gih}%
  \BibitemOpen
  \bibfield  {author} {\bibinfo {author} {\bibfnamefont {R.~C.}\ \bibnamefont
  {Pantig}}, \bibinfo {author} {\bibfnamefont {L.}~\bibnamefont
  {Mastrototaro}}, \bibinfo {author} {\bibfnamefont {G.}~\bibnamefont
  {Lambiase}}, \ and\ \bibinfo {author} {\bibfnamefont {A.}~\bibnamefont
  {\"Ovg\"un}},\ }\bibfield  {title} {\enquote {\bibinfo {title} {{Shadow,
  lensing, quasinormal modes, greybody bounds and neutrino propagation by
  dyonic ModMax black holes}},}\ }\href {\doibase
  10.1140/epjc/s10052-022-11125-y} {\bibfield  {journal} {\bibinfo  {journal}
  {Eur. Phys. J. C}\ }\textbf {\bibinfo {volume} {82}},\ \bibinfo {pages}
  {1155} (\bibinfo {year} {2022}{\natexlab{a}})},\ \Eprint
  {http://arxiv.org/abs/2208.06664}{arXiv:2208.06664 [gr-qc]}\BibitemShut
  {NoStop}%
\bibitem [{\citenamefont {Ayon-Beato}\ and\ \citenamefont
  {Garcia}(1998)}]{Ayon-Beato:1998hmi}%
  \BibitemOpen
  \bibfield  {author} {\bibinfo {author} {\bibfnamefont {E.}~\bibnamefont
  {Ayon-Beato}}\ and\ \bibinfo {author} {\bibfnamefont {A.}~\bibnamefont
  {Garcia}},\ }\bibfield  {title} {\enquote {\bibinfo {title} {{Regular black
  hole in general relativity coupled to nonlinear electrodynamics}},}\ }\href
  {\doibase 10.1103/PhysRevLett.80.5056} {\bibfield  {journal} {\bibinfo
  {journal} {Phys. Rev. Lett.}\ }\textbf {\bibinfo {volume} {80}},\ \bibinfo
  {pages} {5056} (\bibinfo {year} {1998})},\ \Eprint
  {http://arxiv.org/abs/gr-qc/9911046}{arXiv:gr-qc/9911046}\BibitemShut
  {NoStop}%
\bibitem [{\citenamefont {Maeda}\ \emph {et~al.}(2009)\citenamefont {Maeda},
  \citenamefont {Hassaine},\ and\ \citenamefont {Martinez}}]{Maeda:2008ha}%
  \BibitemOpen
  \bibfield  {author} {\bibinfo {author} {\bibfnamefont {H.}~\bibnamefont
  {Maeda}}, \bibinfo {author} {\bibfnamefont {M.}~\bibnamefont {Hassaine}}, \
  and\ \bibinfo {author} {\bibfnamefont {C.}~\bibnamefont {Martinez}},\
  }\bibfield  {title} {\enquote {\bibinfo {title} {{Lovelock black holes with a
  nonlinear Maxwell field}},}\ }\href {\doibase 10.1103/PhysRevD.79.044012}
  {\bibfield  {journal} {\bibinfo  {journal} {Phys. Rev. D}\ }\textbf {\bibinfo
  {volume} {79}},\ \bibinfo {pages} {044012} (\bibinfo {year} {2009})},\
  \Eprint {http://arxiv.org/abs/0812.2038}{arXiv:0812.2038 [gr-qc]}\BibitemShut
  {NoStop}%
\bibitem [{\citenamefont {Mazharimousavi}\ and\ \citenamefont
  {Halilsoy}(2009)}]{Mazharimousavi:2009mb}%
  \BibitemOpen
  \bibfield  {author} {\bibinfo {author} {\bibfnamefont {S.~H.}\ \bibnamefont
  {Mazharimousavi}}\ and\ \bibinfo {author} {\bibfnamefont {M.}~\bibnamefont
  {Halilsoy}},\ }\bibfield  {title} {\enquote {\bibinfo {title} {{Lovelock
  black holes with a power-Yang-Mills source}},}\ }\href {\doibase
  10.1016/j.physletb.2009.10.006} {\bibfield  {journal} {\bibinfo  {journal}
  {Phys. Lett. B}\ }\textbf {\bibinfo {volume} {681}},\ \bibinfo {pages} {190}
  (\bibinfo {year} {2009})},\ \Eprint
  {http://arxiv.org/abs/0908.0308}{arXiv:0908.0308 [gr-qc]}\BibitemShut
  {NoStop}%
\bibitem [{\citenamefont {Habib~Mazharimousavi}\ and\ \citenamefont
  {Halilsoy}(2007)}]{HabibMazharimousavi:2007fst}%
  \BibitemOpen
  \bibfield  {author} {\bibinfo {author} {\bibfnamefont {S.}~\bibnamefont
  {Habib~Mazharimousavi}}\ and\ \bibinfo {author} {\bibfnamefont
  {M.}~\bibnamefont {Halilsoy}},\ }\bibfield  {title} {\enquote {\bibinfo
  {title} {{5D black hole solution in Einstein-Yang-Mills-Gauss-Bonnet
  theory}},}\ }\href {\doibase 10.1103/PhysRevD.76.087501} {\bibfield
  {journal} {\bibinfo  {journal} {Phys. Rev. D}\ }\textbf {\bibinfo {volume}
  {76}},\ \bibinfo {pages} {087501} (\bibinfo {year} {2007})},\ \Eprint
  {http://arxiv.org/abs/0801.1562}{arXiv:0801.1562 [gr-qc]}\BibitemShut
  {NoStop}%
\bibitem [{\citenamefont {Mazharimousavi}\ and\ \citenamefont
  {Halilsoy}(2008)}]{Mazharimousavi:2008ap}%
  \BibitemOpen
  \bibfield  {author} {\bibinfo {author} {\bibfnamefont {S.~H.}\ \bibnamefont
  {Mazharimousavi}}\ and\ \bibinfo {author} {\bibfnamefont {M.}~\bibnamefont
  {Halilsoy}},\ }\bibfield  {title} {\enquote {\bibinfo {title}
  {{Einstein-Yang-Mills black hole solution in higher dimensions by the Wu-Yang
  Ansatz}},}\ }\href {\doibase 10.1016/j.physletb.2007.11.006} {\bibfield
  {journal} {\bibinfo  {journal} {Phys. Lett. B}\ }\textbf {\bibinfo {volume}
  {659}},\ \bibinfo {pages} {471} (\bibinfo {year} {2008})},\ \Eprint
  {http://arxiv.org/abs/0801.1554}{arXiv:0801.1554 [gr-qc]}\BibitemShut
  {NoStop}%
\bibitem [{\citenamefont {Hendi}\ \emph {et~al.}(2018)\citenamefont {Hendi},
  \citenamefont {Panah},\ and\ \citenamefont {Panahiyan}}]{Hendi:2017lgb}%
  \BibitemOpen
  \bibfield  {author} {\bibinfo {author} {\bibfnamefont {S.~H.}\ \bibnamefont
  {Hendi}}, \bibinfo {author} {\bibfnamefont {B.~E.}\ \bibnamefont {Panah}}, \
  and\ \bibinfo {author} {\bibfnamefont {S.}~\bibnamefont {Panahiyan}},\
  }\bibfield  {title} {\enquote {\bibinfo {title} {{Black Hole Solutions in
  Gauss-Bonnet-Massive Gravity in the Presence of Power-Maxwell Field}},}\
  }\href {\doibase 10.1002/prop.201800005} {\bibfield  {journal} {\bibinfo
  {journal} {Fortsch. Phys.}\ }\textbf {\bibinfo {volume} {66}},\ \bibinfo
  {pages} {1800005} (\bibinfo {year} {2018})},\ \Eprint
  {http://arxiv.org/abs/1708.02239}{arXiv:1708.02239 [hep-th]}\BibitemShut
  {NoStop}%
\bibitem [{\citenamefont {Zhang}\ \emph {et~al.}(2015)\citenamefont {Zhang},
  \citenamefont {Yang}, \citenamefont {Zou}, \citenamefont {Xu},\ and\
  \citenamefont {Yue}}]{Zhang:2014eap}%
  \BibitemOpen
  \bibfield  {author} {\bibinfo {author} {\bibfnamefont {M.}~\bibnamefont
  {Zhang}}, \bibinfo {author} {\bibfnamefont {Z.-Y.}\ \bibnamefont {Yang}},
  \bibinfo {author} {\bibfnamefont {D.-C.}\ \bibnamefont {Zou}}, \bibinfo
  {author} {\bibfnamefont {W.}~\bibnamefont {Xu}}, \ and\ \bibinfo {author}
  {\bibfnamefont {R.-H.}\ \bibnamefont {Yue}},\ }\bibfield  {title} {\enquote
  {\bibinfo {title} {{$P-V$ criticality of AdS black hole in the
  Einstein-Maxwell-power-Yang-Mills gravity}},}\ }\href {\doibase
  10.1007/s10714-015-1851-2} {\bibfield  {journal} {\bibinfo  {journal} {Gen.
  Rel. Grav.}\ }\textbf {\bibinfo {volume} {47}},\ \bibinfo {pages} {14}
  (\bibinfo {year} {2015})},\ \Eprint
  {http://arxiv.org/abs/1412.1197}{arXiv:1412.1197 [hep-th]}\BibitemShut
  {NoStop}%
\bibitem [{\citenamefont {Yerra}\ and\ \citenamefont
  {Chandrasekhar}(2019)}]{Yerra:2018mni}%
  \BibitemOpen
  \bibfield  {author} {\bibinfo {author} {\bibfnamefont {P.~K.}\ \bibnamefont
  {Yerra}}\ and\ \bibinfo {author} {\bibfnamefont {B.}~\bibnamefont
  {Chandrasekhar}},\ }\bibfield  {title} {\enquote {\bibinfo {title} {{Heat
  engines at criticality for nonlinearly charged black holes}},}\ }\href
  {\doibase 10.1142/S021773231950216X} {\bibfield  {journal} {\bibinfo
  {journal} {Mod. Phys. Lett. A}\ }\textbf {\bibinfo {volume} {34}},\ \bibinfo
  {pages} {1950216} (\bibinfo {year} {2019})},\ \Eprint
  {http://arxiv.org/abs/1806.08226}{arXiv:1806.08226 [hep-th]}\BibitemShut
  {NoStop}%
\bibitem [{\citenamefont {Hendi}\ \emph
  {et~al.}(2017{\natexlab{a}})\citenamefont {Hendi}, \citenamefont
  {Eslam~Panah}, \citenamefont {Panahiyan},\ and\ \citenamefont
  {Sheykhi}}]{Hendi:2017mgb}%
  \BibitemOpen
  \bibfield  {author} {\bibinfo {author} {\bibfnamefont {S.~H.}\ \bibnamefont
  {Hendi}}, \bibinfo {author} {\bibfnamefont {B.}~\bibnamefont {Eslam~Panah}},
  \bibinfo {author} {\bibfnamefont {S.}~\bibnamefont {Panahiyan}}, \ and\
  \bibinfo {author} {\bibfnamefont {A.}~\bibnamefont {Sheykhi}},\ }\bibfield
  {title} {\enquote {\bibinfo {title} {{Dilatonic BTZ black holes with
  power-law field}},}\ }\href {\doibase 10.1016/j.physletb.2017.01.066}
  {\bibfield  {journal} {\bibinfo  {journal} {Phys. Lett. B}\ }\textbf
  {\bibinfo {volume} {767}},\ \bibinfo {pages} {214} (\bibinfo {year}
  {2017}{\natexlab{a}})},\ \Eprint
  {http://arxiv.org/abs/1703.03403}{arXiv:1703.03403 [gr-qc]}\BibitemShut
  {NoStop}%
\bibitem [{\citenamefont {Hendi}\ \emph
  {et~al.}(2017{\natexlab{b}})\citenamefont {Hendi}, \citenamefont
  {Eslam~Panah}, \citenamefont {Panahiyan},\ and\ \citenamefont
  {Talezadeh}}]{Hendi:2016usw}%
  \BibitemOpen
  \bibfield  {author} {\bibinfo {author} {\bibfnamefont {S.~H.}\ \bibnamefont
  {Hendi}}, \bibinfo {author} {\bibfnamefont {B.}~\bibnamefont {Eslam~Panah}},
  \bibinfo {author} {\bibfnamefont {S.}~\bibnamefont {Panahiyan}}, \ and\
  \bibinfo {author} {\bibfnamefont {M.~S.}\ \bibnamefont {Talezadeh}},\
  }\bibfield  {title} {\enquote {\bibinfo {title} {{Geometrical thermodynamics
  and P-V criticality of black holes with power-law Maxwell field}},}\ }\href
  {\doibase 10.1140/epjc/s10052-017-4693-0} {\bibfield  {journal} {\bibinfo
  {journal} {Eur. Phys. J. C}\ }\textbf {\bibinfo {volume} {77}},\ \bibinfo
  {pages} {133} (\bibinfo {year} {2017}{\natexlab{b}})},\ \Eprint
  {http://arxiv.org/abs/1612.00721}{arXiv:1612.00721 [hep-th]}\BibitemShut
  {NoStop}%
\bibitem [{\citenamefont {Feng}\ \emph {et~al.}(2021)\citenamefont {Feng},
  \citenamefont {Zhou}, \citenamefont {He}, \citenamefont {Zhou},\ and\
  \citenamefont {Yang}}]{Feng:2020swq}%
  \BibitemOpen
  \bibfield  {author} {\bibinfo {author} {\bibfnamefont {Z.-W.}\ \bibnamefont
  {Feng}}, \bibinfo {author} {\bibfnamefont {X.}~\bibnamefont {Zhou}}, \bibinfo
  {author} {\bibfnamefont {G.}~\bibnamefont {He}}, \bibinfo {author}
  {\bibfnamefont {S.-Q.}\ \bibnamefont {Zhou}}, \ and\ \bibinfo {author}
  {\bibfnamefont {S.-Z.}\ \bibnamefont {Yang}},\ }\bibfield  {title} {\enquote
  {\bibinfo {title} {{Joule\textendash{}Thomson expansion of higher dimensional
  nonlinearly AdS black hole with power Maxwell invariant source}},}\ }\href
  {\doibase 10.1088/1572-9494/abecd9} {\bibfield  {journal} {\bibinfo
  {journal} {Commun. Theor. Phys.}\ }\textbf {\bibinfo {volume} {73}},\
  \bibinfo {pages} {065401} (\bibinfo {year} {2021})},\ \Eprint
  {http://arxiv.org/abs/2009.02172}{arXiv:2009.02172 [gr-qc]}\BibitemShut
  {NoStop}%
\bibitem [{\citenamefont {Biswas}(2021)}]{Biswas:2021uop}%
  \BibitemOpen
  \bibfield  {author} {\bibinfo {author} {\bibfnamefont {A.}~\bibnamefont
  {Biswas}},\ }\bibfield  {title} {\enquote {\bibinfo {title} {{Joule-Thomson
  expansion of AdS black holes in Einstein Power-Yang-mills gravity}},}\ }\href
  {\doibase 10.1088/1402-4896/ac2b42} {\bibfield  {journal} {\bibinfo
  {journal} {Phys. Scripta}\ }\textbf {\bibinfo {volume} {96}},\ \bibinfo
  {pages} {125310} (\bibinfo {year} {2021})},\ \Eprint
  {http://arxiv.org/abs/2106.11066}{arXiv:2106.11066 [gr-qc]}\BibitemShut
  {NoStop}%
\bibitem [{\citenamefont {Synge}(1966)}]{Synge:1966okc}%
  \BibitemOpen
  \bibfield  {author} {\bibinfo {author} {\bibfnamefont {J.~L.}\ \bibnamefont
  {Synge}},\ }\bibfield  {title} {\enquote {\bibinfo {title} {{The Escape of
  Photons from Gravitationally Intense Stars}},}\ }\href {\doibase
  10.1093/mnras/131.3.463} {\bibfield  {journal} {\bibinfo  {journal} {Mon.
  Not. Roy. Astron. Soc.}\ }\textbf {\bibinfo {volume} {131}},\ \bibinfo
  {pages} {463} (\bibinfo {year} {1966})}\BibitemShut {NoStop}%
\bibitem [{\citenamefont {Luminet}(1979)}]{Luminet:1979nyg}%
  \BibitemOpen
  \bibfield  {author} {\bibinfo {author} {\bibfnamefont {J.~P.}\ \bibnamefont
  {Luminet}},\ }\bibfield  {title} {\enquote {\bibinfo {title} {{Image of a
  spherical black hole with thin accretion disk}},}\ }\href@noop {} {\bibfield
  {journal} {\bibinfo  {journal} {Astron. Astrophys.}\ }\textbf {\bibinfo
  {volume} {75}},\ \bibinfo {pages} {228} (\bibinfo {year} {1979})}\BibitemShut
  {NoStop}%
\bibitem [{\citenamefont {Falcke}\ \emph {et~al.}(2000)\citenamefont {Falcke},
  \citenamefont {Melia},\ and\ \citenamefont {Agol}}]{Falcke:1999pj}%
  \BibitemOpen
  \bibfield  {author} {\bibinfo {author} {\bibfnamefont {H.}~\bibnamefont
  {Falcke}}, \bibinfo {author} {\bibfnamefont {F.}~\bibnamefont {Melia}}, \
  and\ \bibinfo {author} {\bibfnamefont {E.}~\bibnamefont {Agol}},\ }\bibfield
  {title} {\enquote {\bibinfo {title} {{Viewing the shadow of the black hole at
  the galactic center}},}\ }\href {\doibase 10.1086/312423} {\bibfield
  {journal} {\bibinfo  {journal} {Astrophys. J. Lett.}\ }\textbf {\bibinfo
  {volume} {528}},\ \bibinfo {pages} {L13} (\bibinfo {year} {2000})},\ \Eprint
  {http://arxiv.org/abs/astro-ph/9912263}{arXiv:astro-ph/9912263}\BibitemShut
  {NoStop}%
\bibitem [{\citenamefont {Akiyama}\ \emph {et~al.}(2019)\citenamefont {Akiyama}
  \emph {et~al.}}]{EventHorizonTelescope:2019dse}%
  \BibitemOpen
  \bibfield  {author} {\bibinfo {author} {\bibfnamefont {K.}~\bibnamefont
  {Akiyama}} \emph {et~al.} (\bibinfo {collaboration} {Event Horizon
  Telescope}),\ }\bibfield  {title} {\enquote {\bibinfo {title} {{First M87
  Event Horizon Telescope Results. I. The Shadow of the Supermassive Black
  Hole}},}\ }\href {\doibase 10.3847/2041-8213/ab0ec7} {\bibfield  {journal}
  {\bibinfo  {journal} {Astrophys. J. Lett.}\ }\textbf {\bibinfo {volume}
  {875}},\ \bibinfo {pages} {L1} (\bibinfo {year} {2019})},\ \Eprint
  {http://arxiv.org/abs/1906.11238}{arXiv:1906.11238 [astro-ph.GA]}\BibitemShut
  {NoStop}%
\bibitem [{\citenamefont {Akiyama}\ \emph {et~al.}(2022)\citenamefont {Akiyama}
  \emph {et~al.}}]{EventHorizonTelescope:2022xnr}%
  \BibitemOpen
  \bibfield  {author} {\bibinfo {author} {\bibfnamefont {K.}~\bibnamefont
  {Akiyama}} \emph {et~al.} (\bibinfo {collaboration} {Event Horizon
  Telescope}),\ }\bibfield  {title} {\enquote {\bibinfo {title} {{First
  Sagittarius A* Event Horizon Telescope Results. I. The Shadow of the
  Supermassive Black Hole in the Center of the Milky Way}},}\ }\href {\doibase
  10.3847/2041-8213/ac6674} {\bibfield  {journal} {\bibinfo  {journal}
  {Astrophys. J. Lett.}\ }\textbf {\bibinfo {volume} {930}},\ \bibinfo {pages}
  {L12} (\bibinfo {year} {2022})}\BibitemShut {NoStop}%
\bibitem [{\citenamefont {\"Ovg\"un}\ \emph {et~al.}(2018)\citenamefont
  {\"Ovg\"un}, \citenamefont {Sakall\i{}},\ and\ \citenamefont
  {Saavedra}}]{Ovgun:2018tua}%
  \BibitemOpen
  \bibfield  {author} {\bibinfo {author} {\bibfnamefont {A.}~\bibnamefont
  {\"Ovg\"un}}, \bibinfo {author} {\bibfnamefont {I.}~\bibnamefont
  {Sakall\i{}}}, \ and\ \bibinfo {author} {\bibfnamefont {J.}~\bibnamefont
  {Saavedra}},\ }\bibfield  {title} {\enquote {\bibinfo {title} {{Shadow cast
  and Deflection angle of Kerr-Newman-Kasuya spacetime}},}\ }\href {\doibase
  10.1088/1475-7516/2018/10/041} {\bibfield  {journal} {\bibinfo  {journal}
  {JCAP}\ }\textbf {\bibinfo {volume} {10}},\ \bibinfo {pages} {041} (\bibinfo
  {year} {2018})},\ \Eprint {http://arxiv.org/abs/1807.00388}{arXiv:1807.00388
  [gr-qc]}\BibitemShut {NoStop}%
\bibitem [{\citenamefont {Kocherlakota}\ \emph {et~al.}(2021)\citenamefont
  {Kocherlakota} \emph {et~al.}}]{EventHorizonTelescope:2021dqv}%
  \BibitemOpen
  \bibfield  {author} {\bibinfo {author} {\bibfnamefont {P.}~\bibnamefont
  {Kocherlakota}} \emph {et~al.} (\bibinfo {collaboration} {Event Horizon
  Telescope}),\ }\bibfield  {title} {\enquote {\bibinfo {title} {{Constraints
  on black-hole charges with the 2017 EHT observations of M87*}},}\ }\href
  {\doibase 10.1103/PhysRevD.103.104047} {\bibfield  {journal} {\bibinfo
  {journal} {Phys. Rev. D}\ }\textbf {\bibinfo {volume} {103}},\ \bibinfo
  {pages} {104047} (\bibinfo {year} {2021})},\ \Eprint
  {http://arxiv.org/abs/2105.09343}{arXiv:2105.09343 [gr-qc]}\BibitemShut
  {NoStop}%
\bibitem [{\citenamefont {Belhaj}\ \emph {et~al.}(2021)\citenamefont {Belhaj},
  \citenamefont {Belmahi}, \citenamefont {Benali}, \citenamefont {El~Hadri},
  \citenamefont {El~Moumni},\ and\ \citenamefont
  {Torrente-Lujan}}]{Belhaj:2020okh}%
  \BibitemOpen
  \bibfield  {author} {\bibinfo {author} {\bibfnamefont {A.}~\bibnamefont
  {Belhaj}}, \bibinfo {author} {\bibfnamefont {H.}~\bibnamefont {Belmahi}},
  \bibinfo {author} {\bibfnamefont {M.}~\bibnamefont {Benali}}, \bibinfo
  {author} {\bibfnamefont {W.}~\bibnamefont {El~Hadri}}, \bibinfo {author}
  {\bibfnamefont {H.}~\bibnamefont {El~Moumni}}, \ and\ \bibinfo {author}
  {\bibfnamefont {E.}~\bibnamefont {Torrente-Lujan}},\ }\bibfield  {title}
  {\enquote {\bibinfo {title} {{Shadows of 5D black holes from string
  theory}},}\ }\href {\doibase 10.1016/j.physletb.2020.136025} {\bibfield
  {journal} {\bibinfo  {journal} {Phys. Lett. B}\ }\textbf {\bibinfo {volume}
  {812}},\ \bibinfo {pages} {136025} (\bibinfo {year} {2021})},\ \Eprint
  {http://arxiv.org/abs/2008.13478}{arXiv:2008.13478 [hep-th]}\BibitemShut
  {NoStop}%
\bibitem [{\citenamefont {Belhaj}\ \emph {et~al.}(2020)\citenamefont {Belhaj},
  \citenamefont {Benali}, \citenamefont {El~Balali}, \citenamefont
  {El~Moumni},\ and\ \citenamefont {Ennadifi}}]{Belhaj:2020rdb}%
  \BibitemOpen
  \bibfield  {author} {\bibinfo {author} {\bibfnamefont {A.}~\bibnamefont
  {Belhaj}}, \bibinfo {author} {\bibfnamefont {M.}~\bibnamefont {Benali}},
  \bibinfo {author} {\bibfnamefont {A.}~\bibnamefont {El~Balali}}, \bibinfo
  {author} {\bibfnamefont {H.}~\bibnamefont {El~Moumni}}, \ and\ \bibinfo
  {author} {\bibfnamefont {S.~E.}\ \bibnamefont {Ennadifi}},\ }\bibfield
  {title} {\enquote {\bibinfo {title} {{Deflection angle and shadow behaviors
  of quintessential black holes in arbitrary dimensions}},}\ }\href {\doibase
  10.1088/1361-6382/abbaa9} {\bibfield  {journal} {\bibinfo  {journal} {Class.
  Quant. Grav.}\ }\textbf {\bibinfo {volume} {37}},\ \bibinfo {pages} {215004}
  (\bibinfo {year} {2020})},\ \Eprint
  {http://arxiv.org/abs/2006.01078}{arXiv:2006.01078 [gr-qc]}\BibitemShut
  {NoStop}%
\bibitem [{\citenamefont {Belhaj}\ and\ \citenamefont
  {Sekhmani}(2022)}]{Belhaj:2022kek}%
  \BibitemOpen
  \bibfield  {author} {\bibinfo {author} {\bibfnamefont {A.}~\bibnamefont
  {Belhaj}}\ and\ \bibinfo {author} {\bibfnamefont {Y.}~\bibnamefont
  {Sekhmani}},\ }\bibfield  {title} {\enquote {\bibinfo {title} {{Shadows of
  rotating quintessential black holes in
  Einstein\textendash{}Gauss\textendash{}Bonnet gravity with a cloud of
  strings}},}\ }\href {\doibase 10.1007/s10714-022-02902-x} {\bibfield
  {journal} {\bibinfo  {journal} {Gen. Rel. Grav.}\ }\textbf {\bibinfo {volume}
  {54}},\ \bibinfo {pages} {17} (\bibinfo {year} {2022})}\BibitemShut {NoStop}%
\bibitem [{\citenamefont {Gogoi}\ \emph
  {et~al.}(2023{\natexlab{c}})\citenamefont {Gogoi}, \citenamefont {Sekhmani},
  \citenamefont {Kalita}, \citenamefont {Gogoi},\ and\ \citenamefont
  {Bora}}]{gogoi_joulethomson_2023}%
  \BibitemOpen
  \bibfield  {author} {\bibinfo {author} {\bibfnamefont {D.~J.}\ \bibnamefont
  {Gogoi}}, \bibinfo {author} {\bibfnamefont {Y.}~\bibnamefont {Sekhmani}},
  \bibinfo {author} {\bibfnamefont {D.}~\bibnamefont {Kalita}}, \bibinfo
  {author} {\bibfnamefont {N.~J.}\ \bibnamefont {Gogoi}}, \ and\ \bibinfo
  {author} {\bibfnamefont {J.}~\bibnamefont {Bora}},\ }\bibfield  {title}
  {\enquote {\bibinfo {title} {Joule‐{Thomson} {Expansion} and {Optical}
  {Behaviour} of {Reissner}‐{Nordström}‐{Anti}‐de {Sitter} {Black}
  {Holes} in {Rastall} {Gravity} {Surrounded} by a {Quintessence} {Field}},}\
  }\href {\doibase 10.1002/prop.202300010} {\bibfield  {journal} {\bibinfo
  {journal} {Fortschritte der Physik}\ }\textbf {\bibinfo {volume} {71}},\
  \bibinfo {pages} {2300010} (\bibinfo {year}
  {2023}{\natexlab{c}})}\BibitemShut {NoStop}%
\bibitem [{\citenamefont {\"Ovg\"un}\ and\ \citenamefont
  {Sakall\i{}}(2020)}]{Ovgun:2020gjz}%
  \BibitemOpen
  \bibfield  {author} {\bibinfo {author} {\bibfnamefont {A.}~\bibnamefont
  {\"Ovg\"un}}\ and\ \bibinfo {author} {\bibfnamefont {I.}~\bibnamefont
  {Sakall\i{}}},\ }\bibfield  {title} {\enquote {\bibinfo {title} {{Testing
  generalized Einstein\textendash{}Cartan\textendash{}Kibble\textendash{}Sciama
  gravity using weak deflection angle and shadow cast}},}\ }\href {\doibase
  10.1088/1361-6382/abb579} {\bibfield  {journal} {\bibinfo  {journal} {Class.
  Quant. Grav.}\ }\textbf {\bibinfo {volume} {37}},\ \bibinfo {pages} {225003}
  (\bibinfo {year} {2020})},\ \Eprint
  {http://arxiv.org/abs/2005.00982}{arXiv:2005.00982 [gr-qc]}\BibitemShut
  {NoStop}%
\bibitem [{\citenamefont {\"Ovg\"un}\ \emph {et~al.}(2020)\citenamefont
  {\"Ovg\"un}, \citenamefont {Sakall\i{}}, \citenamefont {Saavedra},\ and\
  \citenamefont {Leiva}}]{Ovgun:2019jdo}%
  \BibitemOpen
  \bibfield  {author} {\bibinfo {author} {\bibfnamefont {A.}~\bibnamefont
  {\"Ovg\"un}}, \bibinfo {author} {\bibfnamefont {I.}~\bibnamefont
  {Sakall\i{}}}, \bibinfo {author} {\bibfnamefont {J.}~\bibnamefont
  {Saavedra}}, \ and\ \bibinfo {author} {\bibfnamefont {C.}~\bibnamefont
  {Leiva}},\ }\bibfield  {title} {\enquote {\bibinfo {title} {{Shadow cast of
  noncommutative black holes in Rastall gravity}},}\ }\href {\doibase
  10.1142/S0217732320501631} {\bibfield  {journal} {\bibinfo  {journal} {Mod.
  Phys. Lett. A}\ }\textbf {\bibinfo {volume} {35}},\ \bibinfo {pages}
  {2050163} (\bibinfo {year} {2020})},\ \Eprint
  {http://arxiv.org/abs/1906.05954}{arXiv:1906.05954 [hep-th]}\BibitemShut
  {NoStop}%
\bibitem [{\citenamefont {Kuang}\ and\ \citenamefont
  {\"Ovg\"un}(2022)}]{Kuang:2022xjp}%
  \BibitemOpen
  \bibfield  {author} {\bibinfo {author} {\bibfnamefont {X.-M.}\ \bibnamefont
  {Kuang}}\ and\ \bibinfo {author} {\bibfnamefont {A.}~\bibnamefont
  {\"Ovg\"un}},\ }\bibfield  {title} {\enquote {\bibinfo {title} {{Strong
  gravitational lensing and shadow constraint from M87* of slowly rotating
  Kerr-like black hole}},}\ }\href@noop {} {\  (\bibinfo {year} {2022})},\
  \Eprint {http://arxiv.org/abs/2205.11003}{arXiv:2205.11003
  [gr-qc]}\BibitemShut {NoStop}%
\bibitem [{\citenamefont {Kumaran}\ and\ \citenamefont
  {\"Ovg\"un}(2022)}]{Kumaran:2022soh}%
  \BibitemOpen
  \bibfield  {author} {\bibinfo {author} {\bibfnamefont {Y.}~\bibnamefont
  {Kumaran}}\ and\ \bibinfo {author} {\bibfnamefont {A.}~\bibnamefont
  {\"Ovg\"un}},\ }\bibfield  {title} {\enquote {\bibinfo {title} {{Deflection
  Angle and Shadow of the Reissner\textendash{}Nordstr\"om Black Hole with
  Higher-Order Magnetic Correction in Einstein-Nonlinear-Maxwell Fields}},}\
  }\href {\doibase 10.3390/sym14102054} {\bibfield  {journal} {\bibinfo
  {journal} {Symmetry}\ }\textbf {\bibinfo {volume} {14}},\ \bibinfo {pages}
  {2054} (\bibinfo {year} {2022})},\ \Eprint
  {http://arxiv.org/abs/2210.00468}{arXiv:2210.00468 [gr-qc]}\BibitemShut
  {NoStop}%
\bibitem [{\citenamefont {Mustafa}\ \emph {et~al.}(2022)\citenamefont
  {Mustafa}, \citenamefont {Atamurotov}, \citenamefont {Hussain}, \citenamefont
  {Shaymatov},\ and\ \citenamefont {\"Ovg\"un}}]{Mustafa:2022xod}%
  \BibitemOpen
  \bibfield  {author} {\bibinfo {author} {\bibfnamefont {G.}~\bibnamefont
  {Mustafa}}, \bibinfo {author} {\bibfnamefont {F.}~\bibnamefont {Atamurotov}},
  \bibinfo {author} {\bibfnamefont {I.}~\bibnamefont {Hussain}}, \bibinfo
  {author} {\bibfnamefont {S.}~\bibnamefont {Shaymatov}}, \ and\ \bibinfo
  {author} {\bibfnamefont {A.}~\bibnamefont {\"Ovg\"un}},\ }\bibfield  {title}
  {\enquote {\bibinfo {title} {{Shadows and gravitational weak lensing by the
  Schwarzschild black hole in the string cloud background with quintessential
  field*}},}\ }\href {\doibase 10.1088/1674-1137/ac917f} {\bibfield  {journal}
  {\bibinfo  {journal} {Chin. Phys. C}\ }\textbf {\bibinfo {volume} {46}},\
  \bibinfo {pages} {125107} (\bibinfo {year} {2022})},\ \Eprint
  {http://arxiv.org/abs/2207.07608}{arXiv:2207.07608 [gr-qc]}\BibitemShut
  {NoStop}%
\bibitem [{\citenamefont {Cimdiker}\ \emph {et~al.}(2021)\citenamefont
  {Cimdiker}, \citenamefont {Demir},\ and\ \citenamefont
  {\"Ovg\"un}}]{Cimdiker:2021cpz}%
  \BibitemOpen
  \bibfield  {author} {\bibinfo {author} {\bibfnamefont {I.}~\bibnamefont
  {Cimdiker}}, \bibinfo {author} {\bibfnamefont {D.}~\bibnamefont {Demir}}, \
  and\ \bibinfo {author} {\bibfnamefont {A.}~\bibnamefont {\"Ovg\"un}},\
  }\bibfield  {title} {\enquote {\bibinfo {title} {{Black hole shadow in
  symmergent gravity}},}\ }\href {\doibase 10.1016/j.dark.2021.100900}
  {\bibfield  {journal} {\bibinfo  {journal} {Phys. Dark Univ.}\ }\textbf
  {\bibinfo {volume} {34}},\ \bibinfo {pages} {100900} (\bibinfo {year}
  {2021})},\ \Eprint {http://arxiv.org/abs/2110.11904}{arXiv:2110.11904
  [gr-qc]}\BibitemShut {NoStop}%
\bibitem [{\citenamefont {Okyay}\ and\ \citenamefont
  {\"Ovg\"un}(2022)}]{Okyay:2021nnh}%
  \BibitemOpen
  \bibfield  {author} {\bibinfo {author} {\bibfnamefont {M.}~\bibnamefont
  {Okyay}}\ and\ \bibinfo {author} {\bibfnamefont {A.}~\bibnamefont
  {\"Ovg\"un}},\ }\bibfield  {title} {\enquote {\bibinfo {title} {{Nonlinear
  electrodynamics effects on the black hole shadow, deflection angle,
  quasinormal modes and greybody factors}},}\ }\href {\doibase
  10.1088/1475-7516/2022/01/009} {\bibfield  {journal} {\bibinfo  {journal}
  {JCAP}\ }\textbf {\bibinfo {volume} {01}},\ \bibinfo {pages} {009} (\bibinfo
  {year} {2022})},\ \Eprint {http://arxiv.org/abs/2108.07766}{arXiv:2108.07766
  [gr-qc]}\BibitemShut {NoStop}%
\bibitem [{\citenamefont {Atamurotov}\ \emph {et~al.}(2023)\citenamefont
  {Atamurotov}, \citenamefont {Hussain}, \citenamefont {Mustafa},\ and\
  \citenamefont {\"Ovg\"un}}]{Atamurotov:2022knb}%
  \BibitemOpen
  \bibfield  {author} {\bibinfo {author} {\bibfnamefont {F.}~\bibnamefont
  {Atamurotov}}, \bibinfo {author} {\bibfnamefont {I.}~\bibnamefont {Hussain}},
  \bibinfo {author} {\bibfnamefont {G.}~\bibnamefont {Mustafa}}, \ and\
  \bibinfo {author} {\bibfnamefont {A.}~\bibnamefont {\"Ovg\"un}},\ }\bibfield
  {title} {\enquote {\bibinfo {title} {{Weak deflection angle and shadow cast
  by the charged-Kiselev black hole with cloud of strings in plasma*}},}\
  }\href {\doibase 10.1088/1674-1137/ac9fbb} {\bibfield  {journal} {\bibinfo
  {journal} {Chin. Phys. C}\ }\textbf {\bibinfo {volume} {47}},\ \bibinfo
  {pages} {025102} (\bibinfo {year} {2023})}\BibitemShut {NoStop}%
\bibitem [{\citenamefont {Pantig}\ \emph {et~al.}(2023)\citenamefont {Pantig},
  \citenamefont {Övgün},\ and\ \citenamefont {Demir}}]{Pantig:2022qak}%
  \BibitemOpen
  \bibfield  {author} {\bibinfo {author} {\bibfnamefont {R.~C.}\ \bibnamefont
  {Pantig}}, \bibinfo {author} {\bibfnamefont {A.}~\bibnamefont {Övgün}}, \
  and\ \bibinfo {author} {\bibfnamefont {D.}~\bibnamefont {Demir}},\ }\bibfield
   {title} {\enquote {\bibinfo {title} {Testing symmergent gravity through the
  shadow image and weak field photon deflection by a rotating black hole using
  the {M87}\$\${\textasciicircum}*\$\$ and {Sgr}. \$\${\textbackslash}hbox
  \{{A}\}{\textasciicircum}*\$\$ results},}\ }\href {\doibase
  10.1140/epjc/s10052-023-11400-6} {\bibfield  {journal} {\bibinfo  {journal}
  {The European Physical Journal C}\ }\textbf {\bibinfo {volume} {83}},\
  \bibinfo {pages} {250} (\bibinfo {year} {2023})}\BibitemShut {NoStop}%
\bibitem [{\citenamefont {Abdikamalov}\ \emph {et~al.}(2019)\citenamefont
  {Abdikamalov}, \citenamefont {Abdujabbarov}, \citenamefont {Ayzenberg},
  \citenamefont {Malafarina}, \citenamefont {Bambi},\ and\ \citenamefont
  {Ahmedov}}]{Abdikamalov:2019ztb}%
  \BibitemOpen
  \bibfield  {author} {\bibinfo {author} {\bibfnamefont {A.~B.}\ \bibnamefont
  {Abdikamalov}}, \bibinfo {author} {\bibfnamefont {A.~A.}\ \bibnamefont
  {Abdujabbarov}}, \bibinfo {author} {\bibfnamefont {D.}~\bibnamefont
  {Ayzenberg}}, \bibinfo {author} {\bibfnamefont {D.}~\bibnamefont
  {Malafarina}}, \bibinfo {author} {\bibfnamefont {C.}~\bibnamefont {Bambi}}, \
  and\ \bibinfo {author} {\bibfnamefont {B.}~\bibnamefont {Ahmedov}},\
  }\bibfield  {title} {\enquote {\bibinfo {title} {{Black hole mimicker hiding
  in the shadow: Optical properties of the $\gamma$ metric}},}\ }\href
  {\doibase 10.1103/PhysRevD.100.024014} {\bibfield  {journal} {\bibinfo
  {journal} {Phys. Rev. D}\ }\textbf {\bibinfo {volume} {100}},\ \bibinfo
  {pages} {024014} (\bibinfo {year} {2019})},\ \Eprint
  {http://arxiv.org/abs/1904.06207}{arXiv:1904.06207 [gr-qc]}\BibitemShut
  {NoStop}%
\bibitem [{\citenamefont {Abdujabbarov}\ \emph {et~al.}(2016)\citenamefont
  {Abdujabbarov}, \citenamefont {Juraev}, \citenamefont {Ahmedov},\ and\
  \citenamefont {Stuchlik}}]{Abdujabbarov:2016efm}%
  \BibitemOpen
  \bibfield  {author} {\bibinfo {author} {\bibfnamefont {A.}~\bibnamefont
  {Abdujabbarov}}, \bibinfo {author} {\bibfnamefont {B.}~\bibnamefont
  {Juraev}}, \bibinfo {author} {\bibfnamefont {B.}~\bibnamefont {Ahmedov}}, \
  and\ \bibinfo {author} {\bibfnamefont {Z.}~\bibnamefont {Stuchlik}},\
  }\bibfield  {title} {\enquote {\bibinfo {title} {{Shadow of rotating wormhole
  in plasma environment}},}\ }\href {\doibase 10.1007/s10509-016-2818-9}
  {\bibfield  {journal} {\bibinfo  {journal} {Astrophys. Space Sci.}\ }\textbf
  {\bibinfo {volume} {361}},\ \bibinfo {pages} {226} (\bibinfo {year}
  {2016})}\BibitemShut {NoStop}%
\bibitem [{\citenamefont {Atamurotov}\ and\ \citenamefont
  {Ahmedov}(2015)}]{Atamurotov:2015nra}%
  \BibitemOpen
  \bibfield  {author} {\bibinfo {author} {\bibfnamefont {F.}~\bibnamefont
  {Atamurotov}}\ and\ \bibinfo {author} {\bibfnamefont {B.}~\bibnamefont
  {Ahmedov}},\ }\bibfield  {title} {\enquote {\bibinfo {title} {{Optical
  properties of black hole in the presence of plasma: shadow}},}\ }\href
  {\doibase 10.1103/PhysRevD.92.084005} {\bibfield  {journal} {\bibinfo
  {journal} {Phys. Rev. D}\ }\textbf {\bibinfo {volume} {92}},\ \bibinfo
  {pages} {084005} (\bibinfo {year} {2015})},\ \Eprint
  {http://arxiv.org/abs/1507.08131}{arXiv:1507.08131 [gr-qc]}\BibitemShut
  {NoStop}%
\bibitem [{\citenamefont {Papnoi}\ \emph {et~al.}(2014)\citenamefont {Papnoi},
  \citenamefont {Atamurotov}, \citenamefont {Ghosh},\ and\ \citenamefont
  {Ahmedov}}]{Papnoi:2014aaa}%
  \BibitemOpen
  \bibfield  {author} {\bibinfo {author} {\bibfnamefont {U.}~\bibnamefont
  {Papnoi}}, \bibinfo {author} {\bibfnamefont {F.}~\bibnamefont {Atamurotov}},
  \bibinfo {author} {\bibfnamefont {S.~G.}\ \bibnamefont {Ghosh}}, \ and\
  \bibinfo {author} {\bibfnamefont {B.}~\bibnamefont {Ahmedov}},\ }\bibfield
  {title} {\enquote {\bibinfo {title} {{Shadow of five-dimensional rotating
  Myers-Perry black hole}},}\ }\href {\doibase 10.1103/PhysRevD.90.024073}
  {\bibfield  {journal} {\bibinfo  {journal} {Phys. Rev. D}\ }\textbf {\bibinfo
  {volume} {90}},\ \bibinfo {pages} {024073} (\bibinfo {year} {2014})},\
  \Eprint {http://arxiv.org/abs/1407.0834}{arXiv:1407.0834 [gr-qc]}\BibitemShut
  {NoStop}%
\bibitem [{\citenamefont {Abdujabbarov}\ \emph {et~al.}(2013)\citenamefont
  {Abdujabbarov}, \citenamefont {Atamurotov}, \citenamefont {Kucukakca},
  \citenamefont {Ahmedov},\ and\ \citenamefont {Camci}}]{Abdujabbarov:2012bn}%
  \BibitemOpen
  \bibfield  {author} {\bibinfo {author} {\bibfnamefont {A.}~\bibnamefont
  {Abdujabbarov}}, \bibinfo {author} {\bibfnamefont {F.}~\bibnamefont
  {Atamurotov}}, \bibinfo {author} {\bibfnamefont {Y.}~\bibnamefont
  {Kucukakca}}, \bibinfo {author} {\bibfnamefont {B.}~\bibnamefont {Ahmedov}},
  \ and\ \bibinfo {author} {\bibfnamefont {U.}~\bibnamefont {Camci}},\
  }\bibfield  {title} {\enquote {\bibinfo {title} {{Shadow of Kerr-Taub-NUT
  black hole}},}\ }\href {\doibase 10.1007/s10509-012-1337-6} {\bibfield
  {journal} {\bibinfo  {journal} {Astrophys. Space Sci.}\ }\textbf {\bibinfo
  {volume} {344}},\ \bibinfo {pages} {429} (\bibinfo {year} {2013})},\ \Eprint
  {http://arxiv.org/abs/1212.4949}{arXiv:1212.4949
  [physics.gen-ph]}\BibitemShut {NoStop}%
\bibitem [{\citenamefont {Atamurotov}\ \emph {et~al.}(2013)\citenamefont
  {Atamurotov}, \citenamefont {Abdujabbarov},\ and\ \citenamefont
  {Ahmedov}}]{Atamurotov:2013sca}%
  \BibitemOpen
  \bibfield  {author} {\bibinfo {author} {\bibfnamefont {F.}~\bibnamefont
  {Atamurotov}}, \bibinfo {author} {\bibfnamefont {A.}~\bibnamefont
  {Abdujabbarov}}, \ and\ \bibinfo {author} {\bibfnamefont {B.}~\bibnamefont
  {Ahmedov}},\ }\bibfield  {title} {\enquote {\bibinfo {title} {{Shadow of
  rotating non-Kerr black hole}},}\ }\href {\doibase
  10.1103/PhysRevD.88.064004} {\bibfield  {journal} {\bibinfo  {journal} {Phys.
  Rev. D}\ }\textbf {\bibinfo {volume} {88}},\ \bibinfo {pages} {064004}
  (\bibinfo {year} {2013})}\BibitemShut {NoStop}%
\bibitem [{\citenamefont {Cunha}\ and\ \citenamefont
  {Herdeiro}(2018)}]{Cunha:2018acu}%
  \BibitemOpen
  \bibfield  {author} {\bibinfo {author} {\bibfnamefont {P.~V.~P.}\
  \bibnamefont {Cunha}}\ and\ \bibinfo {author} {\bibfnamefont {C.~A.~R.}\
  \bibnamefont {Herdeiro}},\ }\bibfield  {title} {\enquote {\bibinfo {title}
  {{Shadows and strong gravitational lensing: a brief review}},}\ }\href
  {\doibase 10.1007/s10714-018-2361-9} {\bibfield  {journal} {\bibinfo
  {journal} {Gen. Rel. Grav.}\ }\textbf {\bibinfo {volume} {50}},\ \bibinfo
  {pages} {42} (\bibinfo {year} {2018})},\ \Eprint
  {http://arxiv.org/abs/1801.00860}{arXiv:1801.00860 [gr-qc]}\BibitemShut
  {NoStop}%
\bibitem [{\citenamefont {Gralla}\ \emph {et~al.}(2019)\citenamefont {Gralla},
  \citenamefont {Holz},\ and\ \citenamefont {Wald}}]{Gralla:2019xty}%
  \BibitemOpen
  \bibfield  {author} {\bibinfo {author} {\bibfnamefont {S.~E.}\ \bibnamefont
  {Gralla}}, \bibinfo {author} {\bibfnamefont {D.~E.}\ \bibnamefont {Holz}}, \
  and\ \bibinfo {author} {\bibfnamefont {R.~M.}\ \bibnamefont {Wald}},\
  }\bibfield  {title} {\enquote {\bibinfo {title} {{Black Hole Shadows, Photon
  Rings, and Lensing Rings}},}\ }\href {\doibase 10.1103/PhysRevD.100.024018}
  {\bibfield  {journal} {\bibinfo  {journal} {Phys. Rev. D}\ }\textbf {\bibinfo
  {volume} {100}},\ \bibinfo {pages} {024018} (\bibinfo {year} {2019})},\
  \Eprint {http://arxiv.org/abs/1906.00873}{arXiv:1906.00873
  [astro-ph.HE]}\BibitemShut {NoStop}%
\bibitem [{\citenamefont {Lobos}\ and\ \citenamefont
  {Pantig}(2022)}]{Lobos:2022jsz}%
  \BibitemOpen
  \bibfield  {author} {\bibinfo {author} {\bibfnamefont {N.~J. L.~S.}\
  \bibnamefont {Lobos}}\ and\ \bibinfo {author} {\bibfnamefont {R.~C.}\
  \bibnamefont {Pantig}},\ }\bibfield  {title} {\enquote {\bibinfo {title}
  {Generalized extended uncertainty principle black holes: Shadow and lensing
  in the macro- and microscopic realms},}\ }\href {\doibase
  10.3390/physics4040084} {\bibfield  {journal} {\bibinfo  {journal} {Physics}\
  }\textbf {\bibinfo {volume} {4}},\ \bibinfo {pages} {1318} (\bibinfo {year}
  {2022})}\BibitemShut {NoStop}%
\bibitem [{\citenamefont {Uniyal}\ \emph {et~al.}(2023)\citenamefont {Uniyal},
  \citenamefont {Chakrabarti}, \citenamefont {Pantig},\ and\ \citenamefont
  {\"Ovg\"un}}]{Uniyal:2023inx}%
  \BibitemOpen
  \bibfield  {author} {\bibinfo {author} {\bibfnamefont {A.}~\bibnamefont
  {Uniyal}}, \bibinfo {author} {\bibfnamefont {S.}~\bibnamefont {Chakrabarti}},
  \bibinfo {author} {\bibfnamefont {R.~C.}\ \bibnamefont {Pantig}}, \ and\
  \bibinfo {author} {\bibfnamefont {A.}~\bibnamefont {\"Ovg\"un}},\ }\bibfield
  {title} {\enquote {\bibinfo {title} {{Nonlinearly charged black holes: Shadow
  and Thin-accretion disk}},}\ }\href@noop {} {\  (\bibinfo {year} {2023})},\
  \Eprint {http://arxiv.org/abs/2303.07174}{arXiv:2303.07174
  [gr-qc]}\BibitemShut {NoStop}%
\bibitem [{\citenamefont {Panotopoulos}\ \emph {et~al.}(2021)\citenamefont
  {Panotopoulos}, \citenamefont {Rinc\'on},\ and\ \citenamefont
  {Lopes}}]{Panotopoulos:2021tkk}%
  \BibitemOpen
  \bibfield  {author} {\bibinfo {author} {\bibfnamefont {G.}~\bibnamefont
  {Panotopoulos}}, \bibinfo {author} {\bibfnamefont {A.}~\bibnamefont
  {Rinc\'on}}, \ and\ \bibinfo {author} {\bibfnamefont {I.}~\bibnamefont
  {Lopes}},\ }\bibfield  {title} {\enquote {\bibinfo {title} {{Orbits of light
  rays in scale-dependent gravity: Exact analytical solutions to the null
  geodesic equations}},}\ }\href {\doibase 10.1103/PhysRevD.103.104040}
  {\bibfield  {journal} {\bibinfo  {journal} {Phys. Rev. D}\ }\textbf {\bibinfo
  {volume} {103}},\ \bibinfo {pages} {104040} (\bibinfo {year} {2021})},\
  \Eprint {http://arxiv.org/abs/2104.13611}{arXiv:2104.13611
  [gr-qc]}\BibitemShut {NoStop}%
\bibitem [{\citenamefont {Panotopoulos}\ and\ \citenamefont
  {Rincon}(2022)}]{Panotopoulos:2022bky}%
  \BibitemOpen
  \bibfield  {author} {\bibinfo {author} {\bibfnamefont {G.}~\bibnamefont
  {Panotopoulos}}\ and\ \bibinfo {author} {\bibfnamefont {A.}~\bibnamefont
  {Rincon}},\ }\bibfield  {title} {\enquote {\bibinfo {title} {{Orbits of light
  rays in $(1+2)$-dimensional Einstein\textendash{}power\textendash{}Maxwell
  gravity: Exact analytical solution to the null geodesic equations}},}\ }\href
  {\doibase 10.1016/j.aop.2022.168947} {\bibfield  {journal} {\bibinfo
  {journal} {Annals Phys.}\ }\textbf {\bibinfo {volume} {443}},\ \bibinfo
  {pages} {168947} (\bibinfo {year} {2022})},\ \Eprint
  {http://arxiv.org/abs/2206.03437}{arXiv:2206.03437 [gr-qc]}\BibitemShut
  {NoStop}%
\bibitem [{\citenamefont {Khodadi}\ and\ \citenamefont
  {Lambiase}(2022)}]{Khodadi:2022pqh}%
  \BibitemOpen
  \bibfield  {author} {\bibinfo {author} {\bibfnamefont {M.}~\bibnamefont
  {Khodadi}}\ and\ \bibinfo {author} {\bibfnamefont {G.}~\bibnamefont
  {Lambiase}},\ }\bibfield  {title} {\enquote {\bibinfo {title} {{Probing
  Lorentz symmetry violation using the first image of Sagittarius A*:
  Constraints on standard-model extension coefficients}},}\ }\href {\doibase
  10.1103/PhysRevD.106.104050} {\bibfield  {journal} {\bibinfo  {journal}
  {Phys. Rev. D}\ }\textbf {\bibinfo {volume} {106}},\ \bibinfo {pages}
  {104050} (\bibinfo {year} {2022})},\ \Eprint
  {http://arxiv.org/abs/2206.08601}{arXiv:2206.08601 [gr-qc]}\BibitemShut
  {NoStop}%
\bibitem [{\citenamefont {Khodadi}\ \emph {et~al.}(2021)\citenamefont
  {Khodadi}, \citenamefont {Lambiase},\ and\ \citenamefont
  {Mota}}]{Khodadi:2021gbc}%
  \BibitemOpen
  \bibfield  {author} {\bibinfo {author} {\bibfnamefont {M.}~\bibnamefont
  {Khodadi}}, \bibinfo {author} {\bibfnamefont {G.}~\bibnamefont {Lambiase}}, \
  and\ \bibinfo {author} {\bibfnamefont {D.~F.}\ \bibnamefont {Mota}},\
  }\bibfield  {title} {\enquote {\bibinfo {title} {{No-hair theorem in the wake
  of Event Horizon Telescope}},}\ }\href {\doibase
  10.1088/1475-7516/2021/09/028} {\bibfield  {journal} {\bibinfo  {journal}
  {JCAP}\ }\textbf {\bibinfo {volume} {09}},\ \bibinfo {pages} {028} (\bibinfo
  {year} {2021})},\ \Eprint {http://arxiv.org/abs/2107.00834}{arXiv:2107.00834
  [gr-qc]}\BibitemShut {NoStop}%
\bibitem [{\citenamefont {Zhao}\ \emph {et~al.}(2023)\citenamefont {Zhao},
  \citenamefont {Cai}, \citenamefont {Das}, \citenamefont {Lambiase},
  \citenamefont {Saridakis},\ and\ \citenamefont {Vagenas}}]{Zhao:2023uam}%
  \BibitemOpen
  \bibfield  {author} {\bibinfo {author} {\bibfnamefont {Y.}~\bibnamefont
  {Zhao}}, \bibinfo {author} {\bibfnamefont {Y.}~\bibnamefont {Cai}}, \bibinfo
  {author} {\bibfnamefont {S.}~\bibnamefont {Das}}, \bibinfo {author}
  {\bibfnamefont {G.}~\bibnamefont {Lambiase}}, \bibinfo {author}
  {\bibfnamefont {E.~N.}\ \bibnamefont {Saridakis}}, \ and\ \bibinfo {author}
  {\bibfnamefont {E.~C.}\ \bibnamefont {Vagenas}},\ }\bibfield  {title}
  {\enquote {\bibinfo {title} {{Quasinormal Modes in Noncommutative
  Schwarzschild black holes}},}\ }\href@noop {} {\  (\bibinfo {year} {2023})},\
  \Eprint {http://arxiv.org/abs/2301.09147}{arXiv:2301.09147
  [gr-qc]}\BibitemShut {NoStop}%
\bibitem [{\citenamefont {Khodadi}\ \emph {et~al.}(2020)\citenamefont
  {Khodadi}, \citenamefont {Allahyari}, \citenamefont {Vagnozzi},\ and\
  \citenamefont {Mota}}]{Khodadi:2020jij}%
  \BibitemOpen
  \bibfield  {author} {\bibinfo {author} {\bibfnamefont {M.}~\bibnamefont
  {Khodadi}}, \bibinfo {author} {\bibfnamefont {A.}~\bibnamefont {Allahyari}},
  \bibinfo {author} {\bibfnamefont {S.}~\bibnamefont {Vagnozzi}}, \ and\
  \bibinfo {author} {\bibfnamefont {D.~F.}\ \bibnamefont {Mota}},\ }\bibfield
  {title} {\enquote {\bibinfo {title} {{Black holes with scalar hair in light
  of the Event Horizon Telescope}},}\ }\href {\doibase
  10.1088/1475-7516/2020/09/026} {\bibfield  {journal} {\bibinfo  {journal}
  {JCAP}\ }\textbf {\bibinfo {volume} {09}},\ \bibinfo {pages} {026} (\bibinfo
  {year} {2020})},\ \Eprint {http://arxiv.org/abs/2005.05992}{arXiv:2005.05992
  [gr-qc]}\BibitemShut {NoStop}%
\bibitem [{\citenamefont {Khodadi}\ and\ \citenamefont
  {Saridakis}(2021)}]{Khodadi:2020gns}%
  \BibitemOpen
  \bibfield  {author} {\bibinfo {author} {\bibfnamefont {M.}~\bibnamefont
  {Khodadi}}\ and\ \bibinfo {author} {\bibfnamefont {E.~N.}\ \bibnamefont
  {Saridakis}},\ }\bibfield  {title} {\enquote {\bibinfo {title}
  {{Einstein-\AE{}ther gravity in the light of event horizon telescope
  observations of M87*}},}\ }\href {\doibase 10.1016/j.dark.2021.100835}
  {\bibfield  {journal} {\bibinfo  {journal} {Phys. Dark Univ.}\ }\textbf
  {\bibinfo {volume} {32}},\ \bibinfo {pages} {100835} (\bibinfo {year}
  {2021})},\ \Eprint {http://arxiv.org/abs/2012.05186}{arXiv:2012.05186
  [gr-qc]}\BibitemShut {NoStop}%
\bibitem [{\citenamefont {Vagnozzi}\ \emph {et~al.}(2023)\citenamefont
  {Vagnozzi} \emph {et~al.}}]{Vagnozzi:2022moj}%
  \BibitemOpen
  \bibfield  {author} {\bibinfo {author} {\bibfnamefont {S.}~\bibnamefont
  {Vagnozzi}} \emph {et~al.},\ }\bibfield  {title} {\enquote {\bibinfo {title}
  {{Horizon-scale tests of gravity theories and fundamental physics from the
  Event Horizon Telescope image of Sagittarius A}},}\ }\href {\doibase
  10.1088/1361-6382/acd97b} {\bibfield  {journal} {\bibinfo  {journal} {Class.
  Quant. Grav.}\ }\textbf {\bibinfo {volume} {40}},\ \bibinfo {pages} {165007}
  (\bibinfo {year} {2023})},\ \Eprint
  {http://arxiv.org/abs/2205.07787}{arXiv:2205.07787 [gr-qc]}\BibitemShut
  {NoStop}%
\bibitem [{\citenamefont {Khodadi}(2022)}]{Khodadi:2022ulo}%
  \BibitemOpen
  \bibfield  {author} {\bibinfo {author} {\bibfnamefont {M.}~\bibnamefont
  {Khodadi}},\ }\bibfield  {title} {\enquote {\bibinfo {title} {{Shadow of
  black hole surrounded by magnetized plasma: Axion-plasmon cloud}},}\ }\href
  {\doibase 10.1016/j.nuclphysb.2022.116014} {\bibfield  {journal} {\bibinfo
  {journal} {Nucl. Phys. B}\ }\textbf {\bibinfo {volume} {985}},\ \bibinfo
  {pages} {116014} (\bibinfo {year} {2022})},\ \Eprint
  {http://arxiv.org/abs/2211.00300}{arXiv:2211.00300 [gr-qc]}\BibitemShut
  {NoStop}%
\bibitem [{\citenamefont {Zhang}\ and\ \citenamefont
  {Guo}(2020)}]{Zhang:2019glo}%
  \BibitemOpen
  \bibfield  {author} {\bibinfo {author} {\bibfnamefont {M.}~\bibnamefont
  {Zhang}}\ and\ \bibinfo {author} {\bibfnamefont {M.}~\bibnamefont {Guo}},\
  }\bibfield  {title} {\enquote {\bibinfo {title} {{Can shadows reflect phase
  structures of black holes?}}}\ }\href {\doibase
  10.1140/epjc/s10052-020-8389-5} {\bibfield  {journal} {\bibinfo  {journal}
  {Eur. Phys. J. C}\ }\textbf {\bibinfo {volume} {80}},\ \bibinfo {pages} {790}
  (\bibinfo {year} {2020})},\ \Eprint
  {http://arxiv.org/abs/1909.07033}{arXiv:1909.07033 [gr-qc]}\BibitemShut
  {NoStop}%
\bibitem [{\citenamefont {Jusufi}(2020)}]{Jusufi:2020dhz}%
  \BibitemOpen
  \bibfield  {author} {\bibinfo {author} {\bibfnamefont {K.}~\bibnamefont
  {Jusufi}},\ }\bibfield  {title} {\enquote {\bibinfo {title} {{Connection
  Between the Shadow Radius and Quasinormal Modes in Rotating Spacetimes}},}\
  }\href {\doibase 10.1103/PhysRevD.101.124063} {\bibfield  {journal} {\bibinfo
   {journal} {Phys. Rev. D}\ }\textbf {\bibinfo {volume} {101}},\ \bibinfo
  {pages} {124063} (\bibinfo {year} {2020})},\ \Eprint
  {http://arxiv.org/abs/2004.04664}{arXiv:2004.04664 [gr-qc]}\BibitemShut
  {NoStop}%
\bibitem [{\citenamefont {Jusufi}(2021)}]{Jusufi:2020mmy}%
  \BibitemOpen
  \bibfield  {author} {\bibinfo {author} {\bibfnamefont {K.}~\bibnamefont
  {Jusufi}},\ }\bibfield  {title} {\enquote {\bibinfo {title} {{Correspondence
  between quasinormal modes and the shadow radius in a wormhole spacetime}},}\
  }\href {\doibase 10.1007/s10714-021-02856-6} {\bibfield  {journal} {\bibinfo
  {journal} {Gen. Rel. Grav.}\ }\textbf {\bibinfo {volume} {53}},\ \bibinfo
  {pages} {87} (\bibinfo {year} {2021})},\ \Eprint
  {http://arxiv.org/abs/2007.16019}{arXiv:2007.16019 [gr-qc]}\BibitemShut
  {NoStop}%
\bibitem [{\citenamefont {Biswas}(2022)}]{Biswas:2022qyl}%
  \BibitemOpen
  \bibfield  {author} {\bibinfo {author} {\bibfnamefont {A.}~\bibnamefont
  {Biswas}},\ }\bibfield  {title} {\enquote {\bibinfo {title} {{Black holes in
  4D AdS Einstein Gauss Bonnet gravity with power: Yang Mills field}},}\ }\href
  {\doibase 10.1007/s10714-022-03047-7} {\bibfield  {journal} {\bibinfo
  {journal} {Gen. Rel. Grav.}\ }\textbf {\bibinfo {volume} {54}},\ \bibinfo
  {pages} {161} (\bibinfo {year} {2022})},\ \Eprint
  {http://arxiv.org/abs/2208.02290}{arXiv:2208.02290 [gr-qc]}\BibitemShut
  {NoStop}%
\bibitem [{\citenamefont {Bouhmadi-L\'opez}\ \emph {et~al.}(2020)\citenamefont
  {Bouhmadi-L\'opez}, \citenamefont {Brahma}, \citenamefont {Chen},
  \citenamefont {Chen},\ and\ \citenamefont {Yeom}}]{Bouhmadi-Lopez:2020oia}%
  \BibitemOpen
  \bibfield  {author} {\bibinfo {author} {\bibfnamefont {M.}~\bibnamefont
  {Bouhmadi-L\'opez}}, \bibinfo {author} {\bibfnamefont {S.}~\bibnamefont
  {Brahma}}, \bibinfo {author} {\bibfnamefont {C.-Y.}\ \bibnamefont {Chen}},
  \bibinfo {author} {\bibfnamefont {P.}~\bibnamefont {Chen}}, \ and\ \bibinfo
  {author} {\bibfnamefont {D.-h.}\ \bibnamefont {Yeom}},\ }\bibfield  {title}
  {\enquote {\bibinfo {title} {{A consistent model of non-singular
  Schwarzschild black hole in loop quantum gravity and its quasinormal
  modes}},}\ }\href {\doibase 10.1088/1475-7516/2020/07/066} {\bibfield
  {journal} {\bibinfo  {journal} {JCAP}\ }\textbf {\bibinfo {volume} {07}},\
  \bibinfo {pages} {066} (\bibinfo {year} {2020})},\ \Eprint
  {http://arxiv.org/abs/2004.13061}{arXiv:2004.13061 [gr-qc]}\BibitemShut
  {NoStop}%
\bibitem [{\citenamefont {Schutz}\ and\ \citenamefont
  {Will}(1985)}]{Schutz:1985km}%
  \BibitemOpen
  \bibfield  {author} {\bibinfo {author} {\bibfnamefont {B.~F.}\ \bibnamefont
  {Schutz}}\ and\ \bibinfo {author} {\bibfnamefont {C.~M.}\ \bibnamefont
  {Will}},\ }\bibfield  {title} {\enquote {\bibinfo {title} {{BLACK HOLE NORMAL
  MODES: A SEMIANALYTIC APPROACH}},}\ }\href {\doibase 10.1086/184453}
  {\bibfield  {journal} {\bibinfo  {journal} {Astrophys. J. Lett.}\ }\textbf
  {\bibinfo {volume} {291}},\ \bibinfo {pages} {L33} (\bibinfo {year}
  {1985})}\BibitemShut {NoStop}%
\bibitem [{\citenamefont {Iyer}\ and\ \citenamefont
  {Will}(1987)}]{Iyer:1986np}%
  \BibitemOpen
  \bibfield  {author} {\bibinfo {author} {\bibfnamefont {S.}~\bibnamefont
  {Iyer}}\ and\ \bibinfo {author} {\bibfnamefont {C.~M.}\ \bibnamefont
  {Will}},\ }\bibfield  {title} {\enquote {\bibinfo {title} {{Black Hole Normal
  Modes: A {WKB} Approach. 1. Foundations and Application of a Higher Order
  {WKB} Analysis of Potential Barrier Scattering}},}\ }\href {\doibase
  10.1103/PhysRevD.35.3621} {\bibfield  {journal} {\bibinfo  {journal} {Phys.
  Rev. D}\ }\textbf {\bibinfo {volume} {35}},\ \bibinfo {pages} {3621}
  (\bibinfo {year} {1987})}\BibitemShut {NoStop}%
\bibitem [{\citenamefont {Konoplya}(2003)}]{Konoplya:2003ii}%
  \BibitemOpen
  \bibfield  {author} {\bibinfo {author} {\bibfnamefont {R.~A.}\ \bibnamefont
  {Konoplya}},\ }\bibfield  {title} {\enquote {\bibinfo {title} {{Quasinormal
  behavior of the d-dimensional Schwarzschild black hole and higher order WKB
  approach}},}\ }\href {\doibase 10.1103/PhysRevD.68.024018} {\bibfield
  {journal} {\bibinfo  {journal} {Phys. Rev. D}\ }\textbf {\bibinfo {volume}
  {68}},\ \bibinfo {pages} {024018} (\bibinfo {year} {2003})},\ \Eprint
  {http://arxiv.org/abs/gr-qc/0303052}{arXiv:gr-qc/0303052}\BibitemShut
  {NoStop}%
\bibitem [{\citenamefont {Matyjasek}\ and\ \citenamefont
  {Telecka}(2019)}]{Matyjasek:2019eeu}%
  \BibitemOpen
  \bibfield  {author} {\bibinfo {author} {\bibfnamefont {J.}~\bibnamefont
  {Matyjasek}}\ and\ \bibinfo {author} {\bibfnamefont {M.}~\bibnamefont
  {Telecka}},\ }\bibfield  {title} {\enquote {\bibinfo {title} {{Quasinormal
  modes of black holes. II. Pad\'e summation of the higher-order WKB terms}},}\
  }\href {\doibase 10.1103/PhysRevD.100.124006} {\bibfield  {journal} {\bibinfo
   {journal} {Phys. Rev. D}\ }\textbf {\bibinfo {volume} {100}},\ \bibinfo
  {pages} {124006} (\bibinfo {year} {2019})},\ \Eprint
  {http://arxiv.org/abs/1908.09389}{arXiv:1908.09389 [gr-qc]}\BibitemShut
  {NoStop}%
\bibitem [{\citenamefont {Konoplya}\ \emph {et~al.}(2019)\citenamefont
  {Konoplya}, \citenamefont {Zhidenko},\ and\ \citenamefont
  {Zinhailo}}]{Konoplya:2019hlu}%
  \BibitemOpen
  \bibfield  {author} {\bibinfo {author} {\bibfnamefont {R.~A.}\ \bibnamefont
  {Konoplya}}, \bibinfo {author} {\bibfnamefont {A.}~\bibnamefont {Zhidenko}},
  \ and\ \bibinfo {author} {\bibfnamefont {A.~F.}\ \bibnamefont {Zinhailo}},\
  }\bibfield  {title} {\enquote {\bibinfo {title} {{Higher order WKB formula
  for quasinormal modes and grey-body factors: recipes for quick and accurate
  calculations}},}\ }\href {\doibase 10.1088/1361-6382/ab2e25} {\bibfield
  {journal} {\bibinfo  {journal} {Class. Quant. Grav.}\ }\textbf {\bibinfo
  {volume} {36}},\ \bibinfo {pages} {155002} (\bibinfo {year} {2019})},\
  \Eprint {http://arxiv.org/abs/1904.10333}{arXiv:1904.10333
  [gr-qc]}\BibitemShut {NoStop}%
\bibitem [{\citenamefont {Gogoi}\ and\ \citenamefont
  {Goswami}(2023)}]{Gogoi:2022ove}%
  \BibitemOpen
  \bibfield  {author} {\bibinfo {author} {\bibfnamefont {D.~J.}\ \bibnamefont
  {Gogoi}}\ and\ \bibinfo {author} {\bibfnamefont {U.~D.}\ \bibnamefont
  {Goswami}},\ }\bibfield  {title} {\enquote {\bibinfo {title} {{Tideless
  traversable wormholes surrounded by cloud of strings in f(R) gravity}},}\
  }\href {\doibase 10.1088/1475-7516/2023/02/027} {\bibfield  {journal}
  {\bibinfo  {journal} {JCAP}\ }\textbf {\bibinfo {volume} {02}},\ \bibinfo
  {pages} {027} (\bibinfo {year} {2023})},\ \Eprint
  {http://arxiv.org/abs/2208.07055}{arXiv:2208.07055 [gr-qc]}\BibitemShut
  {NoStop}%
\bibitem [{\citenamefont {Eslam~Panah}\ \emph {et~al.}(2020)\citenamefont
  {Eslam~Panah}, \citenamefont {Jafarzade},\ and\ \citenamefont
  {Hendi}}]{EslamPanah:2020hoj}%
  \BibitemOpen
  \bibfield  {author} {\bibinfo {author} {\bibfnamefont {B.}~\bibnamefont
  {Eslam~Panah}}, \bibinfo {author} {\bibfnamefont {K.}~\bibnamefont
  {Jafarzade}}, \ and\ \bibinfo {author} {\bibfnamefont {S.~H.}\ \bibnamefont
  {Hendi}},\ }\bibfield  {title} {\enquote {\bibinfo {title} {{Charged 4D
  Einstein-Gauss-Bonnet-AdS black holes: Shadow, energy emission, deflection
  angle and heat engine}},}\ }\href {\doibase 10.1016/j.nuclphysb.2020.115269}
  {\bibfield  {journal} {\bibinfo  {journal} {Nucl. Phys. B}\ }\textbf
  {\bibinfo {volume} {961}},\ \bibinfo {pages} {115269} (\bibinfo {year}
  {2020})},\ \Eprint {http://arxiv.org/abs/2004.04058}{arXiv:2004.04058
  [hep-th]}\BibitemShut {NoStop}%
\bibitem [{\citenamefont {Konoplya}(2019)}]{Konoplya2019}%
  \BibitemOpen
  \bibfield  {author} {\bibinfo {author} {\bibfnamefont {R.~A.}\ \bibnamefont
  {Konoplya}},\ }\bibfield  {title} {\enquote {\bibinfo {title} {{Shadow of a
  black hole surrounded by dark matter}},}\ }\href {\doibase
  10.1016/j.physletb.2019.05.043} {\bibfield  {journal} {\bibinfo  {journal}
  {Phys. Lett. B}\ }\textbf {\bibinfo {volume} {795}},\ \bibinfo {pages} {1}
  (\bibinfo {year} {2019})},\ \Eprint
  {http://arxiv.org/abs/1905.00064}{arXiv:1905.00064 [gr-qc]}\BibitemShut
  {NoStop}%
\bibitem [{\citenamefont {Jafarzade}\ \emph {et~al.}(2021)\citenamefont
  {Jafarzade}, \citenamefont {Kord~Zangeneh},\ and\ \citenamefont
  {Lobo}}]{Jafarzade:2020ova}%
  \BibitemOpen
  \bibfield  {author} {\bibinfo {author} {\bibfnamefont {K.}~\bibnamefont
  {Jafarzade}}, \bibinfo {author} {\bibfnamefont {M.}~\bibnamefont
  {Kord~Zangeneh}}, \ and\ \bibinfo {author} {\bibfnamefont {F.~S.~N.}\
  \bibnamefont {Lobo}},\ }\bibfield  {title} {\enquote {\bibinfo {title}
  {{Shadow, deflection angle and quasinormal modes of Born-Infeld charged black
  holes}},}\ }\href {\doibase 10.1088/1475-7516/2021/04/008} {\bibfield
  {journal} {\bibinfo  {journal} {JCAP}\ }\textbf {\bibinfo {volume} {04}},\
  \bibinfo {pages} {008} (\bibinfo {year} {2021})},\ \Eprint
  {http://arxiv.org/abs/2010.05755}{arXiv:2010.05755 [gr-qc]}\BibitemShut
  {NoStop}%
\bibitem [{\citenamefont {Pantig}\ \emph
  {et~al.}(2022{\natexlab{b}})\citenamefont {Pantig}, \citenamefont {Yu},
  \citenamefont {Rodulfo},\ and\ \citenamefont {Övgün}}]{pantig_shadow_2022}%
  \BibitemOpen
  \bibfield  {author} {\bibinfo {author} {\bibfnamefont {R.~C.}\ \bibnamefont
  {Pantig}}, \bibinfo {author} {\bibfnamefont {P.~K.}\ \bibnamefont {Yu}},
  \bibinfo {author} {\bibfnamefont {E.~T.}\ \bibnamefont {Rodulfo}}, \ and\
  \bibinfo {author} {\bibfnamefont {A.}~\bibnamefont {Övgün}},\ }\bibfield
  {title} {\enquote {\bibinfo {title} {Shadow and weak deflection angle of
  extended uncertainty principle black hole surrounded with dark matter},}\
  }\href {\doibase 10.1016/j.aop.2021.168722} {\bibfield  {journal} {\bibinfo
  {journal} {Annals of Physics}\ }\textbf {\bibinfo {volume} {436}},\ \bibinfo
  {pages} {168722} (\bibinfo {year} {2022}{\natexlab{b}})}\BibitemShut
  {NoStop}%
\bibitem [{\citenamefont {Karmakar}\ \emph {et~al.}(2023)\citenamefont
  {Karmakar}, \citenamefont {Gogoi},\ and\ \citenamefont
  {Goswami}}]{karmakar_thermodynamics_2023}%
  \BibitemOpen
  \bibfield  {author} {\bibinfo {author} {\bibfnamefont {R.}~\bibnamefont
  {Karmakar}}, \bibinfo {author} {\bibfnamefont {D.~J.}\ \bibnamefont {Gogoi}},
  \ and\ \bibinfo {author} {\bibfnamefont {U.~D.}\ \bibnamefont {Goswami}},\
  }\bibfield  {title} {\enquote {\bibinfo {title} {Thermodynamics and shadows
  of {GUP}-corrected black holes with topological defects in {Bumblebee}
  gravity},}\ }\href {\doibase 10.1016/j.dark.2023.101249} {\bibfield
  {journal} {\bibinfo  {journal} {Physics of the Dark Universe}\ }\textbf
  {\bibinfo {volume} {41}},\ \bibinfo {pages} {101249} (\bibinfo {year}
  {2023})}\BibitemShut {NoStop}%
\bibitem [{\citenamefont {Wei}\ and\ \citenamefont {Liu}(2013)}]{Wei:2013kza}%
  \BibitemOpen
  \bibfield  {author} {\bibinfo {author} {\bibfnamefont {S.-W.}\ \bibnamefont
  {Wei}}\ and\ \bibinfo {author} {\bibfnamefont {Y.-X.}\ \bibnamefont {Liu}},\
  }\bibfield  {title} {\enquote {\bibinfo {title} {{Observing the shadow of
  Einstein-Maxwell-Dilaton-Axion black hole}},}\ }\href {\doibase
  10.1088/1475-7516/2013/11/063} {\bibfield  {journal} {\bibinfo  {journal}
  {JCAP}\ }\textbf {\bibinfo {volume} {11}},\ \bibinfo {pages} {063} (\bibinfo
  {year} {2013})},\ \Eprint {http://arxiv.org/abs/1311.4251}{arXiv:1311.4251
  [gr-qc]}\BibitemShut {NoStop}%
\bibitem [{\citenamefont {Kruglov}(2021{\natexlab{a}})}]{Kruglov:2021qzd}%
  \BibitemOpen
  \bibfield  {author} {\bibinfo {author} {\bibfnamefont {S.~I.}\ \bibnamefont
  {Kruglov}},\ }\bibfield  {title} {\enquote {\bibinfo {title}
  {{Einstein\textendash{}Gauss\textendash{}Bonnet Gravity with Nonlinear
  Electrodynamics: Entropy, Energy Emission, Quasinormal Modes and Deflection
  Angle}},}\ }\href {\doibase 10.3390/sym13060944} {\bibfield  {journal}
  {\bibinfo  {journal} {Symmetry}\ }\textbf {\bibinfo {volume} {13}},\ \bibinfo
  {pages} {944} (\bibinfo {year} {2021}{\natexlab{a}})}\BibitemShut {NoStop}%
\bibitem [{\citenamefont {Kruglov}(2021{\natexlab{b}})}]{Kruglov:2021stm}%
  \BibitemOpen
  \bibfield  {author} {\bibinfo {author} {\bibfnamefont {S.~I.}\ \bibnamefont
  {Kruglov}},\ }\bibfield  {title} {\enquote {\bibinfo {title} {{Einstein
  \ensuremath{-} Gauss \ensuremath{-} Bonnet gravity with nonlinear
  electrodynamics}},}\ }\href {\doibase 10.1016/j.aop.2021.168449} {\bibfield
  {journal} {\bibinfo  {journal} {Annals Phys.}\ }\textbf {\bibinfo {volume}
  {428}},\ \bibinfo {pages} {168449} (\bibinfo {year} {2021}{\natexlab{b}})},\
  \Eprint {http://arxiv.org/abs/2104.08099}{arXiv:2104.08099
  [gr-qc]}\BibitemShut {NoStop}%
\bibitem [{\citenamefont {Konoplya}\ and\ \citenamefont
  {Zhidenko}(2011)}]{Konoplya:2011qq}%
  \BibitemOpen
  \bibfield  {author} {\bibinfo {author} {\bibfnamefont {R.~A.}\ \bibnamefont
  {Konoplya}}\ and\ \bibinfo {author} {\bibfnamefont {A.}~\bibnamefont
  {Zhidenko}},\ }\bibfield  {title} {\enquote {\bibinfo {title} {{Quasinormal
  modes of black holes: From astrophysics to string theory}},}\ }\href
  {\doibase 10.1103/RevModPhys.83.793} {\bibfield  {journal} {\bibinfo
  {journal} {Rev. Mod. Phys.}\ }\textbf {\bibinfo {volume} {83}},\ \bibinfo
  {pages} {793} (\bibinfo {year} {2011})},\ \Eprint
  {http://arxiv.org/abs/1102.4014}{arXiv:1102.4014 [gr-qc]}\BibitemShut
  {NoStop}%
\bibitem [{\citenamefont {Cuadros-Melgar}\ \emph {et~al.}(2020)\citenamefont
  {Cuadros-Melgar}, \citenamefont {Fontana},\ and\ \citenamefont
  {de~Oliveira}}]{Cuadros-Melgar:2020kqn}%
  \BibitemOpen
  \bibfield  {author} {\bibinfo {author} {\bibfnamefont {B.}~\bibnamefont
  {Cuadros-Melgar}}, \bibinfo {author} {\bibfnamefont {R.~D.~B.}\ \bibnamefont
  {Fontana}}, \ and\ \bibinfo {author} {\bibfnamefont {J.}~\bibnamefont
  {de~Oliveira}},\ }\bibfield  {title} {\enquote {\bibinfo {title} {{Analytical
  correspondence between shadow radius and black hole quasinormal
  frequencies}},}\ }\href {\doibase 10.1016/j.physletb.2020.135966} {\bibfield
  {journal} {\bibinfo  {journal} {Phys. Lett. B}\ }\textbf {\bibinfo {volume}
  {811}},\ \bibinfo {pages} {135966} (\bibinfo {year} {2020})},\ \Eprint
  {http://arxiv.org/abs/2005.09761}{arXiv:2005.09761 [gr-qc]}\BibitemShut
  {NoStop}%
\end{thebibliography}

%

\end{document}